\documentclass[showpacs,amsmath,amssymb,aps,showkeys,floatfix,prd,a4paper]{revtex4}

\usepackage{dcolumn}
\usepackage{bm}
\usepackage{epsfig}
\usepackage{amsfonts}
\usepackage{amssymb,amscd}

\def\lsim{\raise0.3ex\hbox{$<$\kern-0.75em\raise-1.1ex\hbox{$\sim$}}}

\def\gsim{\raise0.3ex\hbox{$>$\kern-0.75em\raise-1.1ex\hbox{$\sim$}}}

\newcommand{\be}{\begin{equation}}

\newcommand{\ee}{\end{equation}}

\def\beq{\begin{equation}}

\def\eeq{\end{equation}}

\def\beqa{\begin{eqnarray}}

\def\eeqa{\end{eqnarray}}

\newcommand{\ba}{\begin{eqnarray}}

\newcommand{\ea}{\end{eqnarray}}

\def\gappeq{\mathrel{\rlap {\raise.5ex\hbox{$>$}}

{\lower.5ex\hbox{$\sim$}}}}

\def\lappeq{\mathrel{\rlap{\raise.5ex\hbox{$<$}}

{\lower.5ex\hbox{$\sim$}}}}

\def\Toprel#1\over#2{\mathrel{\mathop{#2}\limits^{#1}}}

\begin{document}

\title{Probing a low - mass $Z^{\prime}$ gauge boson at IceCube and prospects for IceCube-Gen2}

\author{Reinaldo {\sc Francener}}
\email{reinaldofrancener@gmail.com}
\affiliation{Instituto de Física Gleb Wataghin - Universidade Estadual de Campinas (UNICAMP), \\ 13083-859, Campinas, SP, Brazil. }

\author{Victor P. {\sc Gon\c{c}alves}}
\email{barros@ufpel.edu.br}
\affiliation{Institute of Physics and Mathematics, Federal University of Pelotas, \\
  Postal Code 354,  96010-900, Pelotas, RS, Brazil}

\author{Diego R. {\sc Gratieri}}
\email{drgratieri@id.uff.br}
\affiliation{Escola de Engenharia Industrial Metal\'urgica de Volta Redonda,
Universidade Federal Fluminense (UFF),\\
 CEP 27255-125, Volta Redonda, RJ, Brazil}
\affiliation{Instituto de Física Gleb Wataghin - Universidade Estadual de Campinas (UNICAMP), \\ 13083-859, Campinas, SP, Brazil. }

\begin{abstract}
In this work, we investigate the impact of the $L_\mu - L_\tau$ model, which predicts a new massive gauge boson, $Z'$, on astrophysical neutrino events at the IceCube Observatory. This new gauge boson couples with leptons from the second and third families, and would break the power law of the astrophysical neutrino flux due to the interaction of this flux with the cosmic neutrino background. We derive the sensitivity of the IceCube to this model considering the HESE data from 12 years of observation by assuming different assumptions for the redshift distributions of astrophysical neutrino sources,  mass ordering, and sum of neutrino masses. Our results indicate that the current IceCube data is able to probe small coupling and masses of the order of some few MeV, with the covered parameter space being larger if a distribution of neutrino sources is described by the star formation rate model. In addition, we demonstrate that the IceCube-Gen2 will cover a large region of the parameter space and will allow us to improve our understanding of the $L_\mu - L_\tau$ model.
\end{abstract}



\maketitle

\vspace{1cm}

\section{Introduction}

One of the main results obtained by the IceCube Collaboration is the discovery and description of a flux of high-energy neutrinos of extragalactic origin \cite{HESE1,HESE2,IceCube:2024fxo,tracksNorte,cascades}. Current data indicate that this flux is approximately isotropic and is often parameterized by a power law of the type $\phi_\nu \propto \Phi_0\, E_{\nu}^{-\gamma}$ \cite{HESE2}. The origin of these neutrinos remains an open question, as only a few of them have been linked to identified point sources by IceCube. The first identified source was the blazar TXS 0506+056 \cite{blazar1,blazar2}, followed by the active galaxy NGC 1068 \cite{sn1068}. With the increase in data collection time at the IceCube Observatory and the expectation of its next generation operating at the beginning of the next decade, the IceCube-Gen2 \cite{IceCube-Gen2:2020qha}, it is becoming possible to search for more than one component of the astrophysical neutrino flux \cite{HESE2} (see also Ref. \cite{Goncalves:2022uwy}), as well as physics effects beyond the standard model, both in the production \cite{IceCube:2018tkk,Berlin:2024lwe} and propagation \cite{Shoemaker:2015qul,Esteban:2021tub,Bustamante:2020mep,Agarwalla:2023sng,KA:2023dyz,Wang:2025qap} of these neutrinos.

One of the simplest scenarios for the beyond the standard model that has a strong impact on the propagation of astrophysical neutrinos is the $L_\mu - L_\tau$ model \cite{He:1990pn,He:1991qd}. This model is based on gauging the difference between the lepton numbers of the second and third generations, and predicts a new massive gauge boson, $Z'$, which couples with the leptons of the second and third families. The interaction Lagrangian associated with this theory beyond the standard model is given by
\begin{eqnarray}
    \mathcal{L} = g' Z_{\alpha}' [
    (\bar{\mu} \gamma^{\alpha} \mu) - 
    (\bar{\tau} \gamma^{\alpha} \tau) + 
    (\bar{\nu}_{\mu} \gamma^{\alpha} P_{L} \nu_{\mu}) - 
    (\bar{\nu}_{\tau} \gamma^{\alpha} P_{L} \nu_{\tau})
    ] ,
    \label{eq:lagrangian}
\end{eqnarray}
where $g'$ is the coupling of the new boson. One of the main motivations of the $L_\mu - L_\tau$ model is that it appears as a minimalist alternative to explain the discrepancy between the theoretical and experimentally observed anomalous magnetic moment of the muon $(g-2)_\mu$ \cite{Keshavarzi:2018mgv}.

Recent works have shown that IceCube can be sensitive to variations in the flux of astrophysical neutrinos caused by a $Z'$ boson with a mass of a few MeV \cite{Kamada:2015era,Araki:2015mya,DiFranzo:2015qea,Carpio:2021jhu,Hooper:2023fqn,delaVega:2024pbk}.These previous analyses used different IceCube data: 988 days of data collection from all high-energy event topologies \cite{Kamada:2015era,Araki:2015mya,DiFranzo:2015qea} and six years of high-energy cascade data \cite{Carpio:2021jhu,Hooper:2023fqn,delaVega:2024pbk}. The variations in the flux would be caused by the annihilation of high-energy neutrinos with the cosmic neutrino background, giving rise to the $Z'$ boson, which subsequently decays into two neutrinos, regenerating the flux in a region less energetic than the initial astrophysical neutrino. One of the goals of this paper is to update this previous studies by using the  High Energy Start Events (HESE) data from 12 years of data collection \cite{IceCube:2023sov},  searching for the parameters of $Z'$ mass and coupling that improve the description of the data, as well as the regions that can be excluded from the parameter space. Our second goal is to investigate, for the first time, the sensitivity of the future IceCube-Gen2 Observatory to the presence of a $Z'$ gauge boson described by the $L_\mu - L_\tau$ model. In our analysis, we will present results for the sensitivity considering different assumptions for the redshift distributions of astrophysical neutrino sources, mass ordering, and sum of neutrino masses.

Our paper is organized as follows: in the next section, we present a brief review of the formalism used to calculate the propagation of the astrophysical neutrino flux, considering the neutrino - neutrino interaction predicted by the $L_\mu - L_\tau$ model.
Furthermore, we discuss the calculation of events at IceCube and the cross-section of neutrinos with the observatory. In section \ref{sec:res} we present our results for the astrophysical neutrino flux considering the $Z'$ boson, and for the sensitivity of IceCube and IceCube-Gen2 Observatories to this new boson. 
Finally, in section \ref{sec:sum}, we summarize our main results and conclusions.

\section{Formalism}
\label{sec:for}

\subsection{Neutrino propagation}

The spectrum of astrophysical neutrinos, $\phi_i$, for the mass eigenstate $i$ reaching Earth is obtained by solving the set of coupled differential equations given by \cite{Bhattacharjee:1999mup,Hooper:2023fqn}
\begin{eqnarray}
-(1+z)\frac{H(z)}{c}\frac{\mathrm{d}\phi_i}{\mathrm{d}z} = 
J_i(E_0,z)
- \phi_i\sum_{j} \langle n_{\nu_i}(z)\sigma_{ij}(E_0,z) \rangle
+ P_i\int_{E_0}^{\infty} \mathrm{d}E'\sum_{j,k} \phi_k 
\left\langle
n_{\nu_j}(z)\frac{\mathrm{d}\sigma_{kj}}{\mathrm{d}E_0}(E',z)
\right\rangle \, ,
    \label{eq:spectrum}
\end{eqnarray}
where $P_i = \sum_{l} \mathrm{Br}(Z'\rightarrow \nu_l \nu_i)$, $n_{\nu_j}$ is the cosmic neutrino background density of mass eigenstate $j$ and $E_0$ is the neutrino energy measured at Earth. The neutrino energy at the source is expressed in terms of $E_0$ as $E_\nu = E_0 (1+z)$. In our analysis, we will use $n_{\nu} = 336$ per cm$^{3}$, equally distributed between neutrino and antineutrino, and between the three flavors. $H(z)$ is the Hubble rate and in the range of redshift $z$ of our interest here it can be estimated considering the Universe dominated by dark matter and energy, which implies
\begin{eqnarray}
    H(z) \simeq H_0 
    \sqrt{\Omega_\Lambda + \Omega_M(1+z)^{3}}\, ,
    \label{eq:hubble}
\end{eqnarray}
where $H_0$ is the current value of the Hubble rate, and we are approximating $\Omega_\Lambda \approx 1-\Omega_M$. We will also assume the parameters obtained by the best fit of the Planck Collaboration to these quantities \cite{Planck:2018vyg} ($H_0 = 67.4$ km s$^{-1}$ Mpc$^{-1}$ and $\Omega_M = 0.315$). Moreover, $J_i(E_0,z)$ in Eq. (\ref{eq:spectrum}) is the astrophysical neutrino production term.  Usually $J_i$ is parameterized with the product of a power law for the neutrino spectrum and a redshift distribution $f(z)$ of the neutrino sources, as follows:
\begin{eqnarray}
    J_i(E_0,z) \propto \Phi_0 
    E_0^{-\gamma} f(z)\, ,
    \label{eq:J}
\end{eqnarray}
where $\Phi_0$ and $\gamma$ are the flux normalization and spectral index, respectively. In our work, we will consider two different parameterizations for the distribution of neutrino sources. In the first parameterization, we will assume a distribution according to the star formation rate (SFR) model \cite{Yuksel:2008cu}, where the redshift distribution has a maximum at $z \approx 1$. In the second case, we will assume the parameterization derived considering the distribution of BL Lacertae (BLL) objects \cite{Ajello:2013lka}, which consists of a subpopulation of blazars. 
This distribution is concentrated at lower redshifts, 
with a maximum at 0.06.

The last two terms in Eq. (\ref{eq:spectrum}) contribute when there are interactions between the astrophysical neutrinos and cosmic neutrino background. In particular, the penultimate term characterizes the absorption of the flux by annihilation into a $Z'$ boson, while the last one describes the regeneration by the subsequent decay of $Z'$ into neutrinos. The cross-section for the annihilation of a neutrino - antineutrino pair into a $Z'$ boson described by the $L_\mu - L_\tau$ model, considering neutrinos with eigenstates of mass $i$ and $j$, is given by \cite{Hooper:2023fqn}
\begin{eqnarray}
    \sigma_{ij} = 
    \sigma(\nu_i\bar{\nu}_j \rightarrow \nu\bar{\nu}) = 
    \frac{2g'^{4}s(U_{\mu i}^{\dagger}U_{\mu j} - U_{\tau i}^{\dagger}U_{\tau j})^{2}}{3\pi [(s-m_{Z'}^{2})^{2}+m_{Z'}^{2}\Gamma_{Z'}^{2}]}\; ,
    \label{eq:cs}
\end{eqnarray}
where $U_{\alpha i}$ is the Pontecorvo-Maki-Nakagawa-Sakata matrix (PMNS), and $s = 2E_\nu m_{j}$, with $m_j$ being the target neutrino mass. For the PMNS matrix elements, we are using the fit from NuFIT-6.0 \cite{Esteban:2024eli}. Previous studies \cite{Kamada:2015era,Araki:2015mya,DiFranzo:2015qea,Hooper:2023fqn,delaVega:2024pbk} have demonstrated that the IceCube Observatory is sensitive to a low-mass $Z'$ ($\approx 1 - 10$ MeV). In this case, the boson produced does not have enough mass to produce muon pairs in its decay, so it only decays into a neutrino - antineutrino pair, with a decay width $\Gamma_{Z'} = g'^{2 }m_{Z'}/12\pi$.

The thermal average in Eq. (\ref{eq:spectrum}) represents,
\begin{eqnarray}
    \langle n_{\nu_i}(z)\sigma_{ij}(E_0,z) \rangle = 
    \int \frac{\mathrm{d}^3\vec{p}}{(2\pi)^{3}}
    \frac{\sigma(E_0, z, \vec{p})}{\mathrm{e}^{|\vec{p}|/T_0(1+z)}+1} \; ,
    \label{eq:termalAverage}
\end{eqnarray}
with $T_0 \approx 1.95$ K being the temperature of the cosmic neutrino background when $z=0$. The cross-section $\sigma(E_0, z, \vec{p})$ is calculated using Eq. (\ref{eq:cs}), but considering a target neutrino with momentum $\vec{p}$. In the case where the neutrino target mass is much bigger than its momentum, we can assume that $\langle n_{\nu_i}(z)\sigma_{ij}(E_0,z) \rangle = n_{\nu_j} \sigma_{ij}$. In our analysis, we will assume this approximation, except when the lightest neutrino is massless. { Currently, it is possible to obtain the difference between the values of different neutrino mass eigenstates using the data that probes the neutrino oscillation mechanism. For these quantities, we will use the best fit values derived in the Ref.   \cite{Esteban:2024eli}. Regarding the sum of the masses of the three eigenstates, 
we will consider two possibilities: (a) that the lightest neutrino mass eigenstate is massless. Such case will be named inferior hereafter; and (b) that the sum of the masses of the three mass eigenstates is 0.12 eV, which is the maximum value allowed by Planck Collaboration observations at the 95\% confidence level \cite{Planck:2018vyg}. Such case will be denoted superior.}

Finally, once produced, the $Z'$ boson decays into two energetic neutrinos. The differential cross-section $\frac{\mathrm{d}\sigma_{kj}}{\mathrm{d}E_0}(E',z)$ in Eq. (\ref{eq:spectrum}) gives us the distribution of the production of a neutrino of energy $E_0$ from the decay of a boson $Z'$ produced by a neutrino of energy $E'$. This differential cross-section can be estimated by modifying Eq. (\ref{eq:cs}) as follows \cite{Hooper:2023fqn}
\begin{eqnarray}
    \frac{\mathrm{d}\sigma_{kj}}{\mathrm{d}E_0}(E',z) = 
    \sigma_{kj} (E',z) f(E', E_0) \; ,
    \label{eq:distCS}
\end{eqnarray}
where
\begin{eqnarray}
    f(E',E_0) = \frac{3}{E'}
    \left[
    \left( \frac{E_0}{E'} \right)^{2}
    + \left( 1-\frac{E_0}{E'} \right)^{2}
    \right] \Theta(E' - E_0) \; .
    \label{eq:f}
\end{eqnarray}
Such an approximation will be used in our calculations.

\subsection{Neutrino events at IceCube Observatory}

The astrophysical neutrino flux is one of the key ingredients for estimating high-energy neutrino events at observatories like IceCube, which is given by
\begin{eqnarray}
\mathrm{d}N_{events} = 
T \sum_{\alpha, l}
N_{effe,\alpha}(E_{\nu})\times 
\phi_{\nu_l} (E_{\nu})\times
\sigma _{\nu _l \alpha}(E_\nu) \times
T_{\nu_l}(E_\nu , \theta_{z})\,
{\mathrm{d}(E_{vis})\, \mathrm{d}\Omega}\, ,
\label{eq:eventos}
\end{eqnarray}
where $T$ is the observatory exposure time for data collection; $N_{effe,\alpha}$ is the effective number of targets of type $\alpha$ in the observatory;  $\phi_{\nu_l}(E_\nu)$ the neutrino flux described in the previous subsection; $\sigma _{\nu _l \alpha}(E_\nu)$ is the cross-section of neutrinos of flavor $l =e, \mu, \tau$ with targets of type $\alpha$ and $T_{\nu_l}(E_\nu , \theta_z)$ is the transmission coefficient on Earth of neutrinos of flavor $l$ and energy $E_\nu$ that cross the Earth with zenith angle $\theta_z$. The effective number of targets in the observatory is obtained from the estimated effective mass for HESE \cite{ntargets}. For the neutrino transmission coefficient on Earth we are using the results obtained in \cite{Francener:2024rjw}, which considers the Earth's density profile given by PREM and takes into account the effects of flux regeneration by Neutral Current (NC)  neutrino interactions and tau decay produced in Charged Current (CC) neutrino interactions.

An important quantity present in Eq. (\ref{eq:eventos}) is the neutrino - target cross-section (For a recent review, see Ref. \cite{Reno:2023sdm}). For high-energy neutrinos, the most common scattering is the Deep Inelastic Scattering (DIS). This scattering is characterized by the exchange of a  $W^{\pm}$ ($Z^{0}$) boson between the neutrino with four-momentum $k$ and the nucleon with four-momentum $p$ in CC (NC) interactions. In DIS, the exchanged boson has a four-momentum $q = k - k'$, and we consider $-q^{2} = Q^{2} > 1\,\mathrm{GeV}^{2}$, with $k' $ being the four-momentum of the leptonic final state. Moreover, in the interaction, the target nucleon fragments and becomes a hadronic final state with invariant mass $W > 1.4$ GeV. It is common to write this process in terms of the Bjorken variable, $x$, and the inelasticity, $y$, defined with $x = Q^{2}/2p\cdot q$ and $y = p\cdot q /p\cdot k$. In the nucleon rest frame $y$ becomes the fraction of the initial neutrino energy transferred to the target nucleon. The double differential cross-section of the charged current DIS with respect to $x$ and $y$ is given by \cite{paschos,reno}
\begin{eqnarray}
\begin{aligned}
\frac{\mathrm{d}\sigma^{\nu_{l}N}}{\mathrm{d}x\mathrm{d}y}=
\frac{G_F^{2}m_N E_\nu}{\pi}
\left(
\frac{M_W^2}{Q^2+M_W^2}
\right)^2
\left\{
\left(
y^2x+\frac{m_l^2y}{2E_\nu m_N}
\right)F^{CC}_1(x,Q^2)+\right. 
\left(
1-y-\frac{m_l^2 y}{4E_\nu^2}-\frac{m_Nxy}{2E_\nu}
\right)F^{CC}_2(x,Q^2)+ \\
 + \left(
xy-\frac{xy^2}{2}-\frac{m_l^2y}{4E_\nu m_N}
\right)F^{CC}_3(x,Q^2)+ 
\left. \frac{m_l^{2} (m_l^{2}+Q^2)}{E_\nu^2 m_N^2 x}F^{CC}_4(x,Q^2)
-\frac{m_l^2}{E_\nu m_N}F^{CC}_5(x,Q^2)
\right\} \, ,
\end{aligned}
\label{eq:csDIS}
\end{eqnarray}
where $m_l$ is the mass of the charged lepton produced, $M_W$ the mass of the $W^{\pm}$ boson  and $F_{i}^{CC}$ are the nucleon structure functions for the charged current interaction. Such structure functions are expressed in terms of parton distribution functions (PDF's). In our analysis, we will use  the PDFs parameterized by the CT14 collaboration \cite{ct14}. In the case of neutral current interaction, an equation similar to Eq. (\ref{eq:csDIS}) can be derived, differing only in the masses of the final lepton and the exchanged boson, in addition to the structure functions. In order to obtain the total cross-section, we have to integrate Eq. (\ref{eq:csDIS}) with respect to $x$ and $y$ in the kinematically allowed ranges \cite{reno}. 

In addition to DIS interactions with the nucleon, we will consider the interaction of electronic antineutrinos with electrons for the IceCube events. This process, known as Glashow resonance \cite{Glashow:1960zz}, consists in the annihilation of the antineutrino with the electron into an on-shell $W^{-}$ boson, which decays and can be measured through its decay products. The cross-section of this process is dominant for neutrinos with energy close to $M_{W}^{2}/2 m_e$. For more details, we refer the interested reader to the Ref. \cite{Goncalves:2022uwy}.

\section{Results}
\label{sec:res}

\begin{figure}[t]
	\centering
	\begin{tabular}{ccc}
	\includegraphics[width=0.33\textwidth]{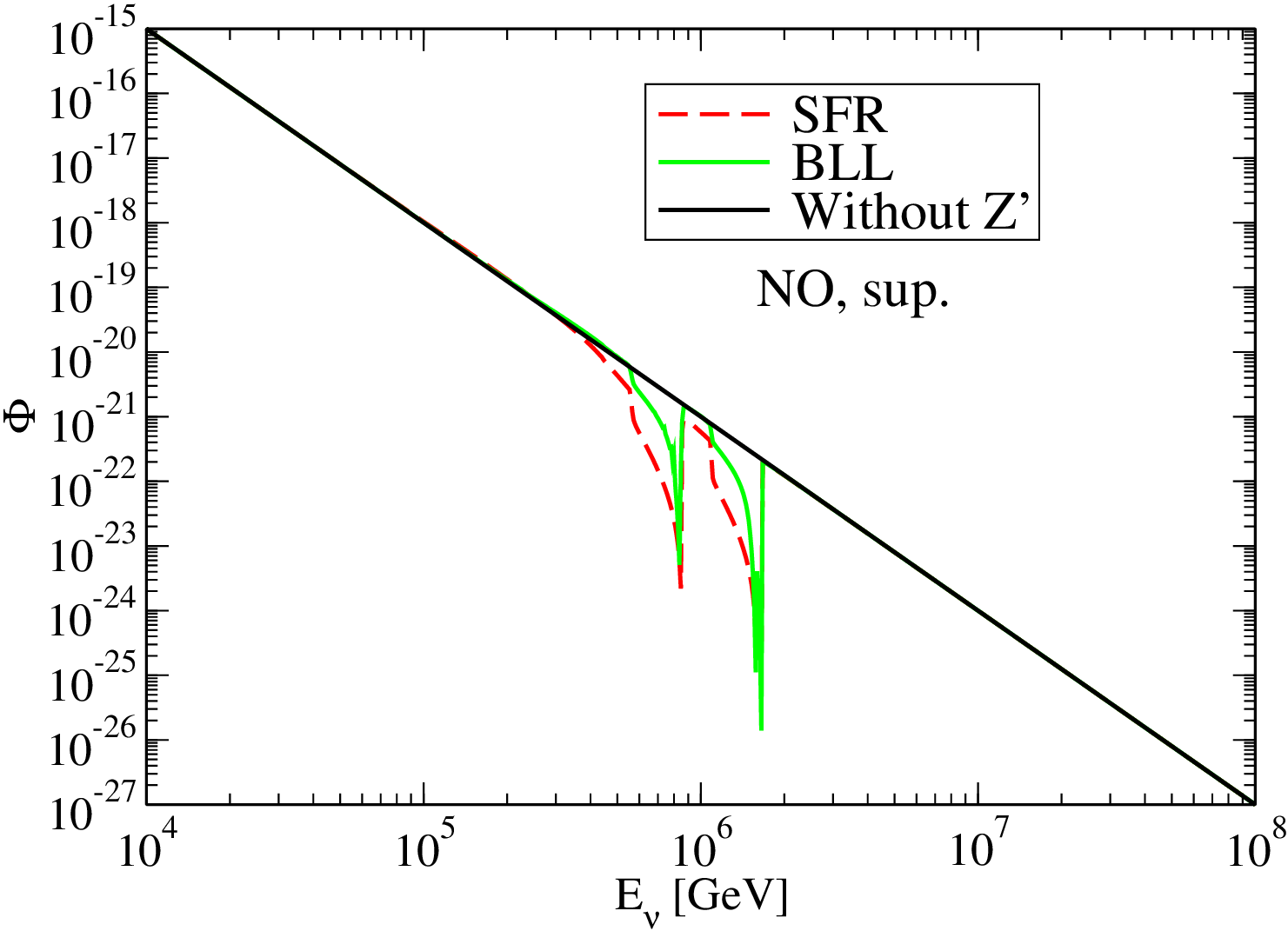}
	\includegraphics[width=0.33\textwidth]{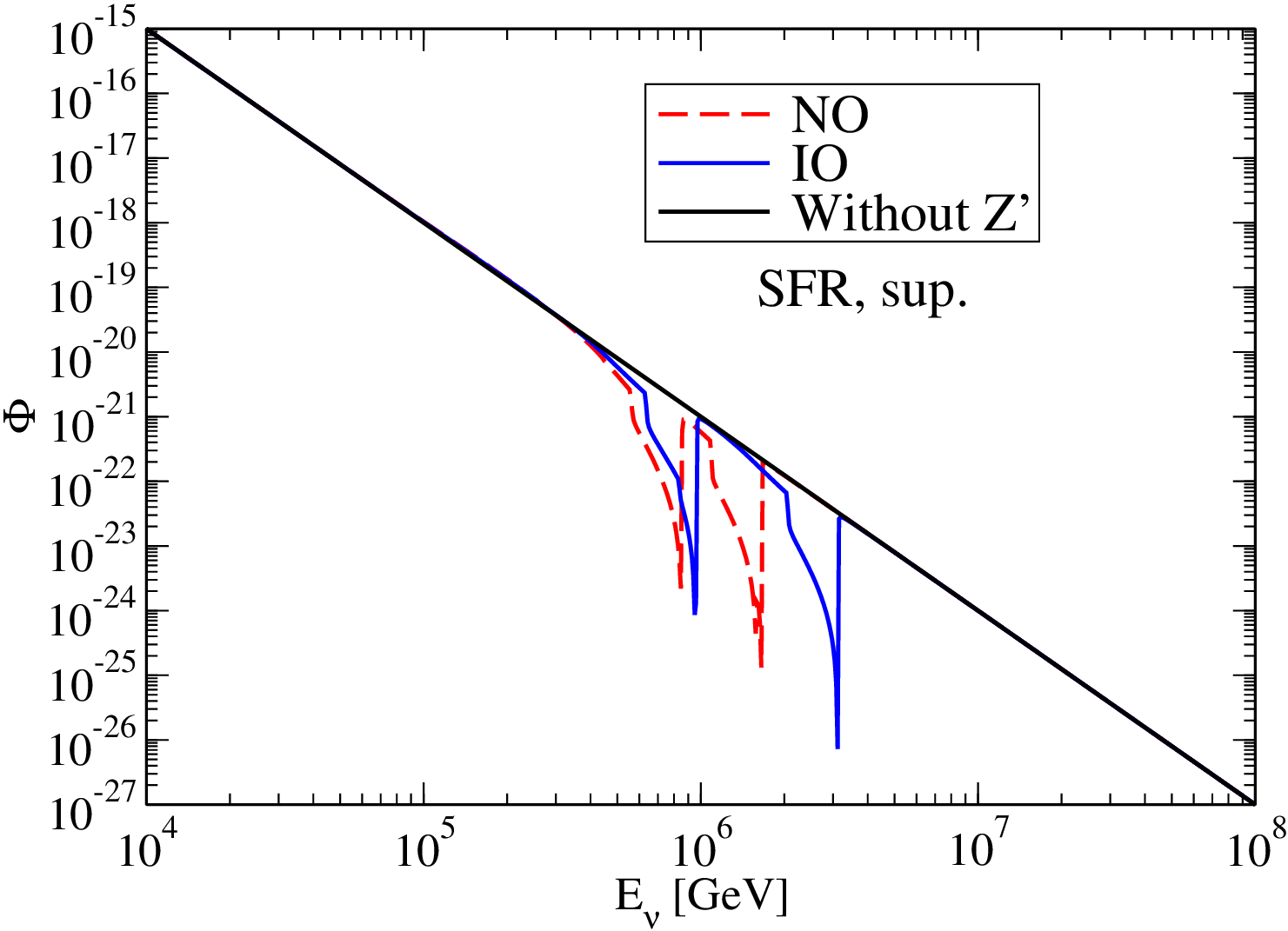} 
	\includegraphics[width=0.33\textwidth]{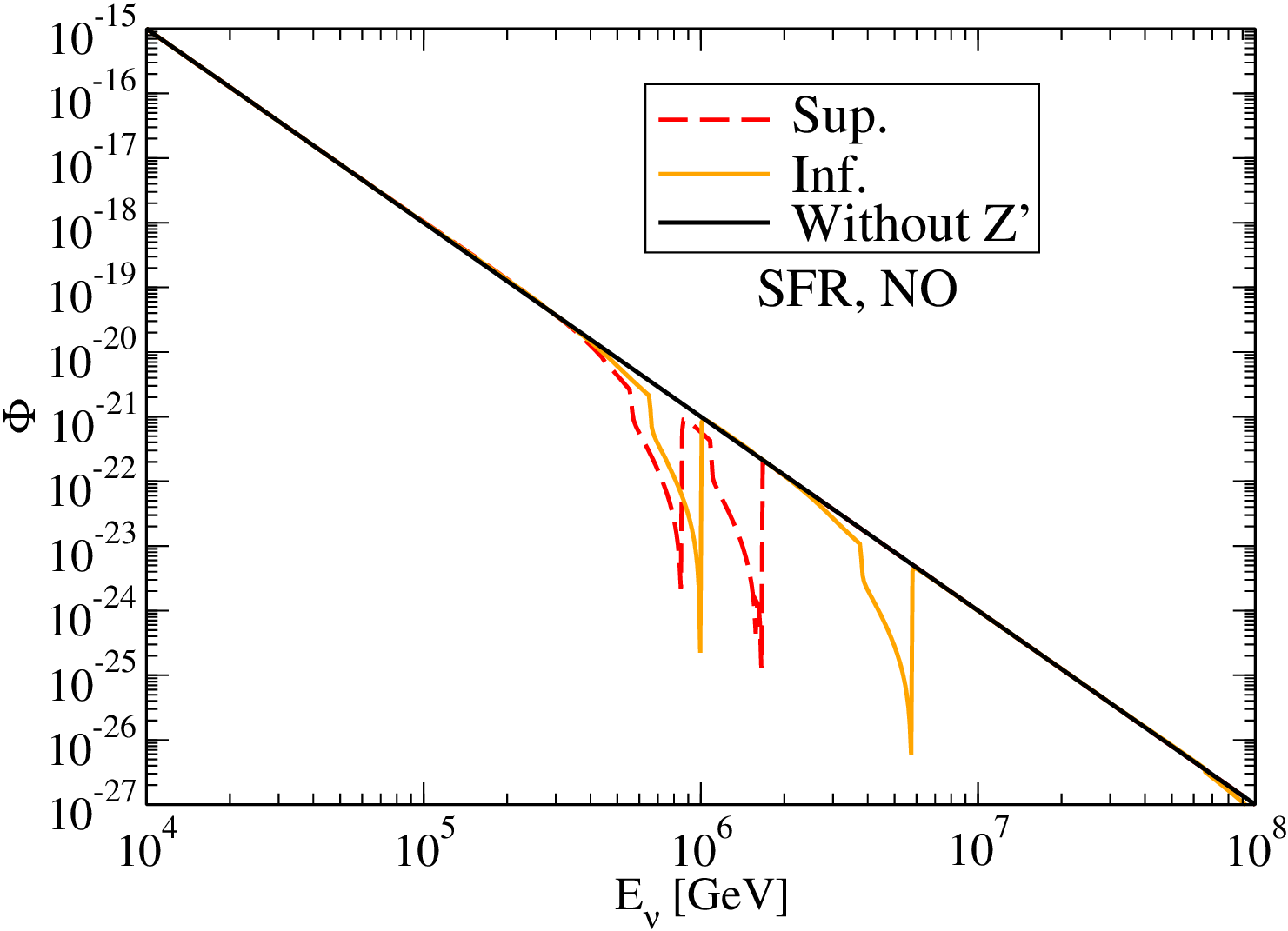} 
			\end{tabular}
\caption{ Neutrino flux as a function of neutrino energy, considering a $Z'$ boson with $m_{Z'} = 10$ MeV and spectral index $\gamma = 3.0$. {\bf Left:} Comparison between SFR and BLL redshift distributions for neutrino sources, assuming Normal Ordering and upper mass limit. {\bf Middle}: Comparison between Normal Ordering (NO) and Inverted Ordering (IO), assuming SFR as the neutrino source and upper mass limit. {\bf Right:} Comparison between upper limit and lower limit for mass sum, assuming SFR as the neutrino source and Normal Ordering. }
\label{fig:flux}
\end{figure}

In order to investigate the impact of the $Z'$ boson in the IceCube events, we will first estimate the impact of this boson on the neutrino flux. In Figure \ref{fig:flux} we present the impact of a $Z'$ boson with $m_{Z'} = 10$ MeV on the flux of astrophysical neutrinos considering different scenarios for the redshift distributions of astrophysical neutrino sources,  mass ordering, and sum of neutrino masses, and comparing with what is expected by the Standard Model. In this analysis, we assume that the flux normalization $\Phi_0$ and the spectral index $\gamma$  are equal to 1.0 and 3.0, respectively. In the left panel of Figure \ref{fig:flux} we present the comparison of the solution of Eq. (\ref{eq:spectrum}), derived considering the redshift distributions SFR and BLL for neutrino sources, and assuming a normal ordering and the superior model for the sum of masses. One has that SFR distribution is characterized by  a redshift distribution concentrated at higher values of $z$, which implies that neutrinos propagate over greater distances and have greater deviations in their energies. As a consequence, one has a larger absorption of neutrinos using this source in comparison to the predictions derived using the BLL  redshift distribution. 
In the middle panel of Fig. \ref{fig:flux}, we compare the flux predictions for normal ordering and inverted ordering of neutrino masses, derived assuming the SFR redshift distribution and the upper limit of the sum of neutrino masses. It is important to emphasize that in the normal ordering case, the two lowest mass eigenvalues are closest, while in inverted one, the closest eigenvalues are the largest eigenvalues. One has that the resulting fluxes are similar, except for the relative distance between dips, which is greater in the inverted ordering. Finally, in the right panel of Fig. \ref{fig:flux} we compare the neutrino flux solutions for the two possibilities for the sum of the neutrino masses, derived assuming the SFR distribution and the normal ordering.  When we assume the lightest neutrino massless, the relative distance of the masses becomes greater, further separating the dips in the flux. The effect of the lower mass neutrino is not visible in the figure for the inferior limit of the neutrino mass sum, as it is found at $E_\nu > 10^{8}$ GeV.

\begin{figure}[t]
	\centering
	\begin{tabular}{ccc}
	\includegraphics[width=0.5\textwidth]{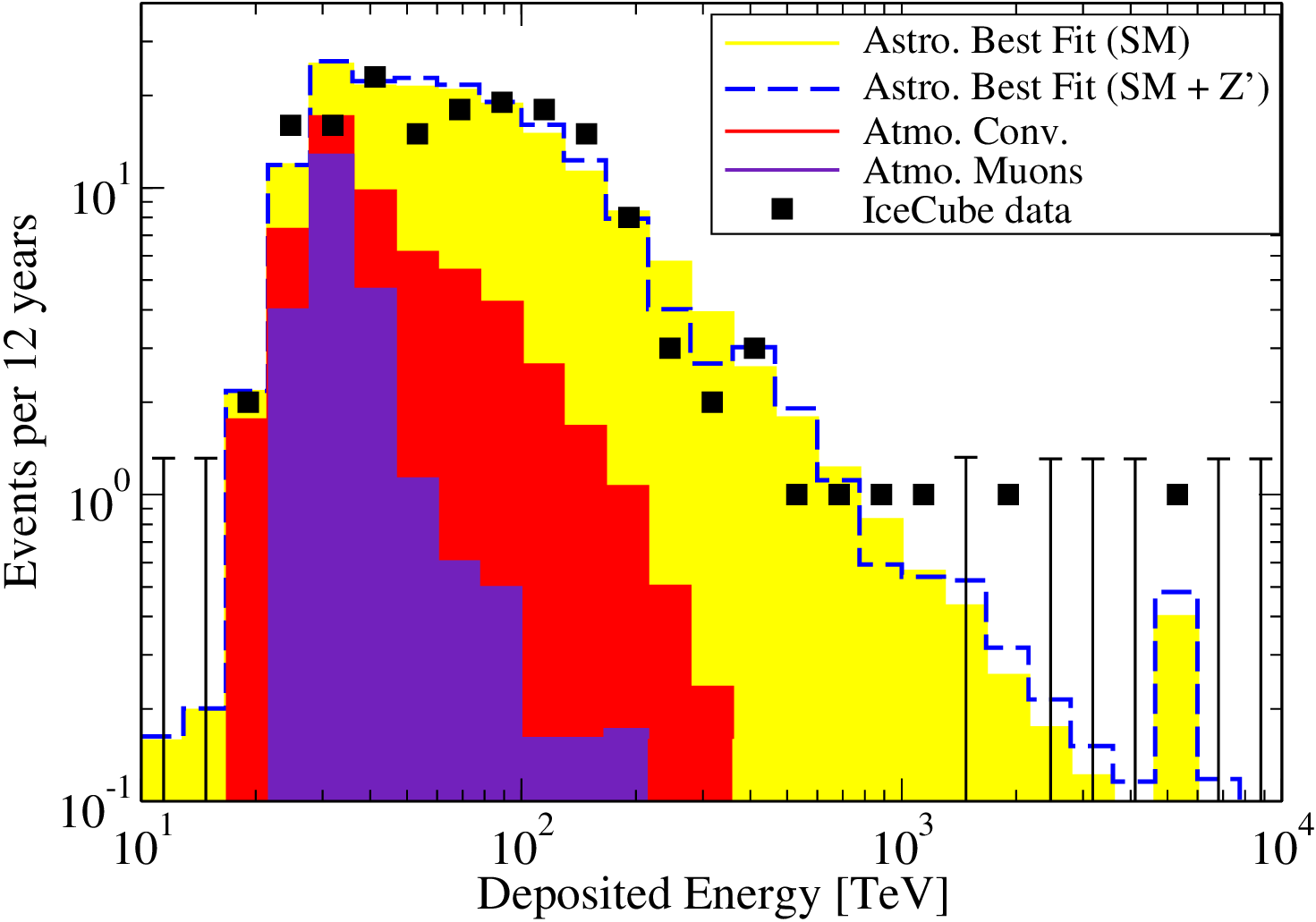}
	\includegraphics[width=0.5\textwidth]{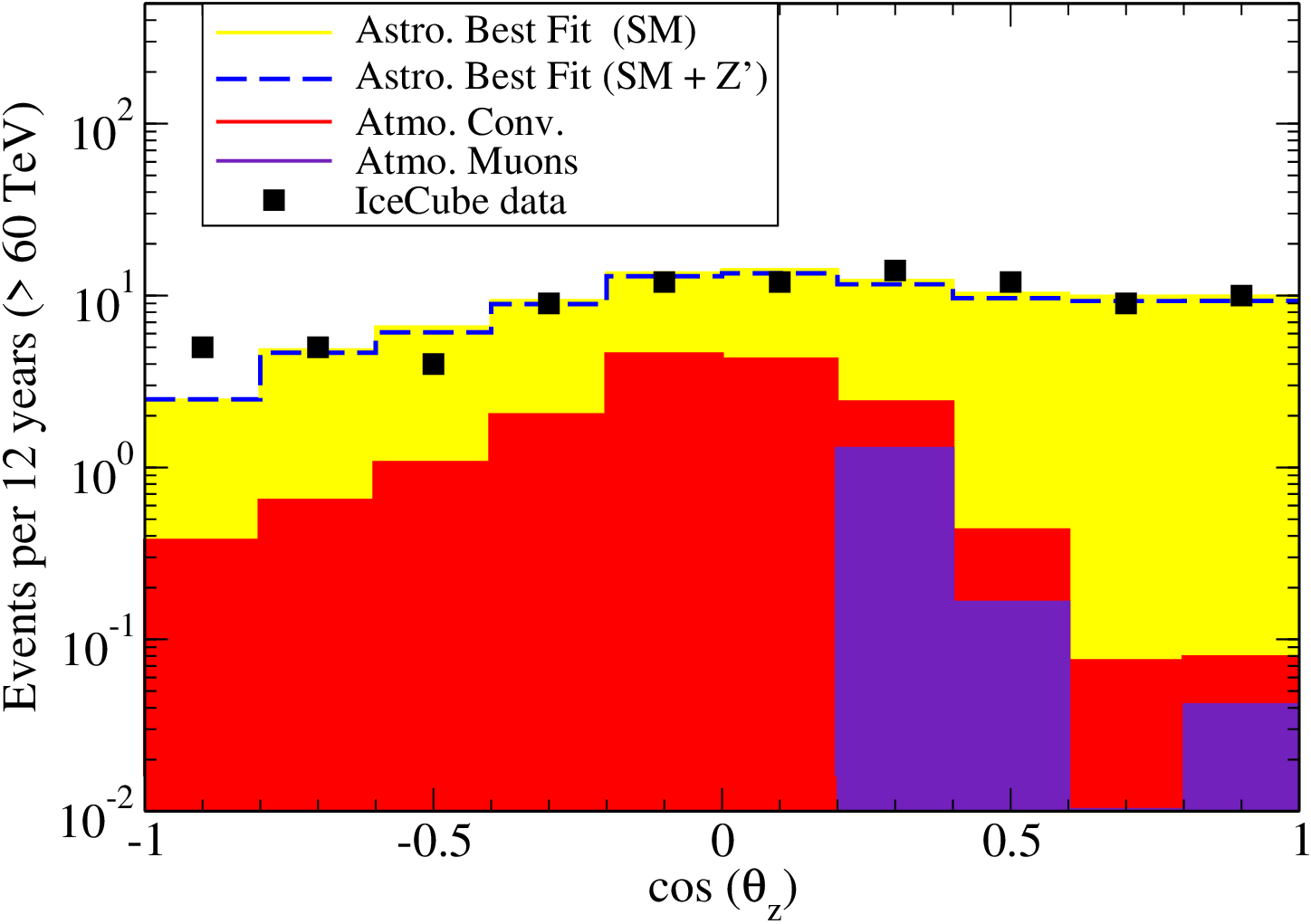} 
			\end{tabular}
\caption{ Number of events in the HESE at IceCube Observatory as a function of deposited energy (left) and azimuthal angle (right) per 12 years. Each color of the histogram specify different contributions to the events, accordingly to the best fit parameters: astrophysical neutrinos (yellow), atmospheric neutrinos (red) and atmospheric muons (purple). In the blue dashed line, we present our results for the best fit obtained considering a $Z'$ boson with mass of 6 MeV, Inverted Ordering, BLL for the redshift distribution and superior limit for the mass sum. }
\label{fig:events}
\end{figure}

Our main goal in this work is to verify the impact of the $L_\mu - L_\tau$ model on the flux of astrophysical neutrinos and, consequently, on the number of HESE events in IceCube. In Fig. \ref{fig:events} we present the distributions of the number of these events in the deposited energy and in the zenith angle for HESE in 12 years of observation. To obtain the best fit parameters of the flux with and without the $Z'$ boson, we will assume the Poisson distribution for the events \cite{Poisson:1837}, which implies the likelihood function is given by
\begin{eqnarray}
L(\vec{\theta}) = \prod_{i}^{N}\frac{\mu_{i}^{n_{i}}\mathrm{e}^{-\mu_{i}}}{n_{i}!}
    \label{eq:likelihood}
\end{eqnarray}
where $\vec{\theta}$ is the set of parameters we seek to maximize the likelihood, $\mu_i$ is the number of expected events and $n_i$ the number of events observed in bin $i$. Using the Neyman-Pearson lemma \cite{Neyman:1933wgr} for the statistical test, $\lambda = -2\,\mathrm{ln}[L(\vec{\theta}_1)/L(\vec{\theta}_2)]$, the best fit of the parameters $\vec{\theta}$ will be the one that minimizes $\lambda$, which is given by
\begin{eqnarray}
\lambda = 2\sum_{i}^{N}\left[ \mu_{i} - n_{i} + n_{i}\,\mathrm{ln} \left( \frac{n_{i}}{\mu_{i}} \right) \right] + \sum_{j}^{m} \left( \frac{c_{j} - c_{j}^{*}}{\sigma_{j}} \right)^{2}\, .
    \label{eq:chi2}
\end{eqnarray}
The last term of the equation above refers to the Pull method \cite{Fogli:2002pt}, used here to adjust the parameters $c_i$ with expectation $c_{i}^{*}$, which include the renormalization of the backgrounds of atmospheric neutrinos and muons and an uncertainty in the energy deposited in each event. We assume the same allowed intervals for $c_i$ and $\sigma_j$ used in \cite{IceCube:2015gsk}. Finally, we assume the validity of Wilks' theorem, which implies that $\lambda$ follows a $\chi^{2}$ distribution \cite{Wilks:1938dza}.

Initially, let's consider the SM case, i.e.  without the inclusion of $Z^{\prime}$ gauge boson. Assuming the methodology described above, we obtain that the best fit of the astrophysical flux is $\Phi_0 = 1.33$ and $\gamma = 2.93$, whose values are similar to those obtained by the IceCube collaboration for the analysis of 7.5 years of data collection \cite{HESE2}. In our analysis, we only have used events with deposited energy above 60 TeV, since in these bins the expected background events are smaller than the number of events expected from astrophysical neutrinos. The distributions of events in the deposited energy and in the zenith angle are presented in Fig. \ref{fig:events}, where the yellow part of the histogram represents astrophysical neutrino events, and in red and purple the expected backgrounds for atmospheric neutrinos and muons, respectively. The Pull method did not change the normalization of the backgrounds, but changed the normalization of the deposited energy with a factor of 1.04. The $\chi^{2}_{min}$ obtained was 9.54.

\begin{table}[b]
\begin{center}
\begin{tabular}{|c|c|c|c|c|c|c|c|}
\hline
Configuration & $m_{Z'}$ (MeV) &$ {\cal{X}}^{2}_{min}$ & $\gamma$ & $\Phi_{0}$ & $k$ & $AIC$ &  $AIC_{c}$   \\
\hline
\hline
Standard Model & -- &  9.54 & 2.93 & 1.33 & 2 & 13.54 & 13.99 \\
\hline
\hline
 NO, inf, BLL & 6 & 7.51 & 2.90 & 1.44 & 4 & 15.51& 17.03  \\
\hline
NO, inf, SFR  & 6 & 8.06 & 2.90 & 1.66  & 4 & 16.06 & 17.66  \\
\hline
NO, sup, BLL  & 7 & 7.43 & 2.80 & 1.40 & 4 & 15.43 &  17.11 \\
\hline
NO, sup, SFR  & 7 & 8.54 & 2.80 & 1.61 & 4 & 16.54 &  18.14  \\
\hline
IO, inf, BLL  & 6 & 7.75 & 2.90 & 1.50 & 4 & 15.75 &  17.35 \\
\hline
IO, inf, SFR  & 9  & 8.57 & 2.80 & 1.28 & 4 & 16.57 &  18.17 \\
\hline
IO, sup, BLL  & 6 & 6.76 & 2.90 & 1.43 &  4 & 14.76 &   16.36 \\
\hline
IO, sup, SFR  & 6 & 7.96 & 2.80 & 1.60 &  4 & 15.96 & 17.56  \\
\hline
\end{tabular}
\caption{Best fit parameters obtained for the IceCube data assuming a $Z'$  boson. We considered different combinations of redshift distribution, mass ordering and limit of masses. We apply the AIC and the AICc criterion  to each model considering the histograms presented in Fig. \ref{fig:events}. The SM result is also presented for comparison. }
\label{table:parameters}
\end{center}
\end{table}

Let's  now consider the presence of a $Z^{\prime}$ gauge boson and different assumptions for the  redshift distributions of astrophysical neutrino sources, mass ordering, and sum of neutrino masses. The associated results for the flux parameters derived using the statistical analysis described above are presented in  Table \ref{table:parameters}. In particular, we present  the $Z'$ masses that improve the description of the current IceCube data, together with the resulting $\chi^ {2}_{min}$ obtained and the flux parameters. The best description of the data occurs for a $Z'$ with a mass of approximately 6 MeV, with $\Delta \chi^{2}$ of 2.03 (2.78) for NO, inf, BLL (IO, sup, BLL), respectively, compared to $\chi^{2}_{min}$ obtained without $Z'$. 
 Such a result agree with those derived in Refs. \cite{Carpio:2021jhu,Hooper:2023fqn,delaVega:2024pbk} using 6 years of IceCube cascade collection data, which also indicated that the $\chi^{2}$ is decreased when a $Z'$ is inserted with masses between approximately 7 and 16 MeV \cite{Carpio:2021jhu}, 5 and 15 MeV \cite{Hooper:2023fqn}, and 6 and 8 MeV \cite{delaVega:2024pbk} for $g' > 10^{-4}$. Our results indicated that the region between 6 and 9 MeV remains favored with 12 years of IceCube HESE data. 
 However, it is important to emphasize that this decrease in $\chi^2$ does not necessarily indicate an improvement in the goodness of fit, since we are adding two new parameters in the analysis.
In order to perform a quantitative comparison of the models we will apply the Akaike Information Criterion (AIC), which rewards goodness of fit (as assessed by the likelihood function), but that also includes a penalty associated with the increasing of the number of estimated parameters. 
The AIC criterion is defined as \cite{Akaike}
\begin{equation}
\text{AIC} = 2k - 2\ln({\cal L}_{max}) = 2k + {\cal X}^{2}_{min},
\label{Eq:AIC}
\end{equation}
where $ k $ is the number of parameters in the model, and $ {\cal L}_{max} $ is the maximum likelihood (the ${\cal{X}}^{2}_{min}$). Furthermore, given the scarce number of counties in each bin on the histograms of Fig. \ref{fig:events}, the AIC criterion must take in account the low statistics and must be corrected as 
\begin{equation}
AIC_{c} = AIC + \frac{2k(k+1)}{n-k-1} ,
\label{Eq:AICc}
\end{equation}
where $n$ is the number of independent data points used to compute the likelihood ({\it i. e.}, the sum of the number of bins in each histogram).  Then it is possible to construct the estimator 
\begin{equation}
\Delta AIC_{c} = AIC^{Z'}_{c} - AIC^{SM}_{c} \,\,.
\label{Eq:Delta_AICc}
\end{equation}
Following Ref. \cite{Burnham}, one has that if $0 \le \Delta AIC_{c} \le 2$ the model is strongly supported, while for $2 < \Delta AIC_{c} \le 4$ the model still is plausible but with less support. Finally, if $\Delta AIC_{c} > 4$ the considered model is unlikely to  be the best one. 
In Table \ref{table:parameters} we also present our results the $AIC$ and $AIC_{c}$ quantities for the SM case as well as for the other models considered. The case "IO, sup, BLL" results a minimum value of the ${\cal X}^{2}_{min}$ among all the models (including the SM prediction) and in the minimum value of $AIC^{Z'}_{c}$ concerning the  $Z'$ hypothesis. However,  in comparison to the SM prediction, we obtain $\Delta AIC_{c} = 2.37$, which means that this case could be nearly as good as the SM, but not a better option than the standard model. As a consequence, one has that the improvement of the ${\cal X}^2_{min}$ is not enough to compensate the two extra parameters, and that a stronger conclusion about the presence of a $Z'$ gauge boson is still not possible.

\begin{figure}[t]
	\centering
	\begin{tabular}{ccc}
	\includegraphics[width=0.5\textwidth]{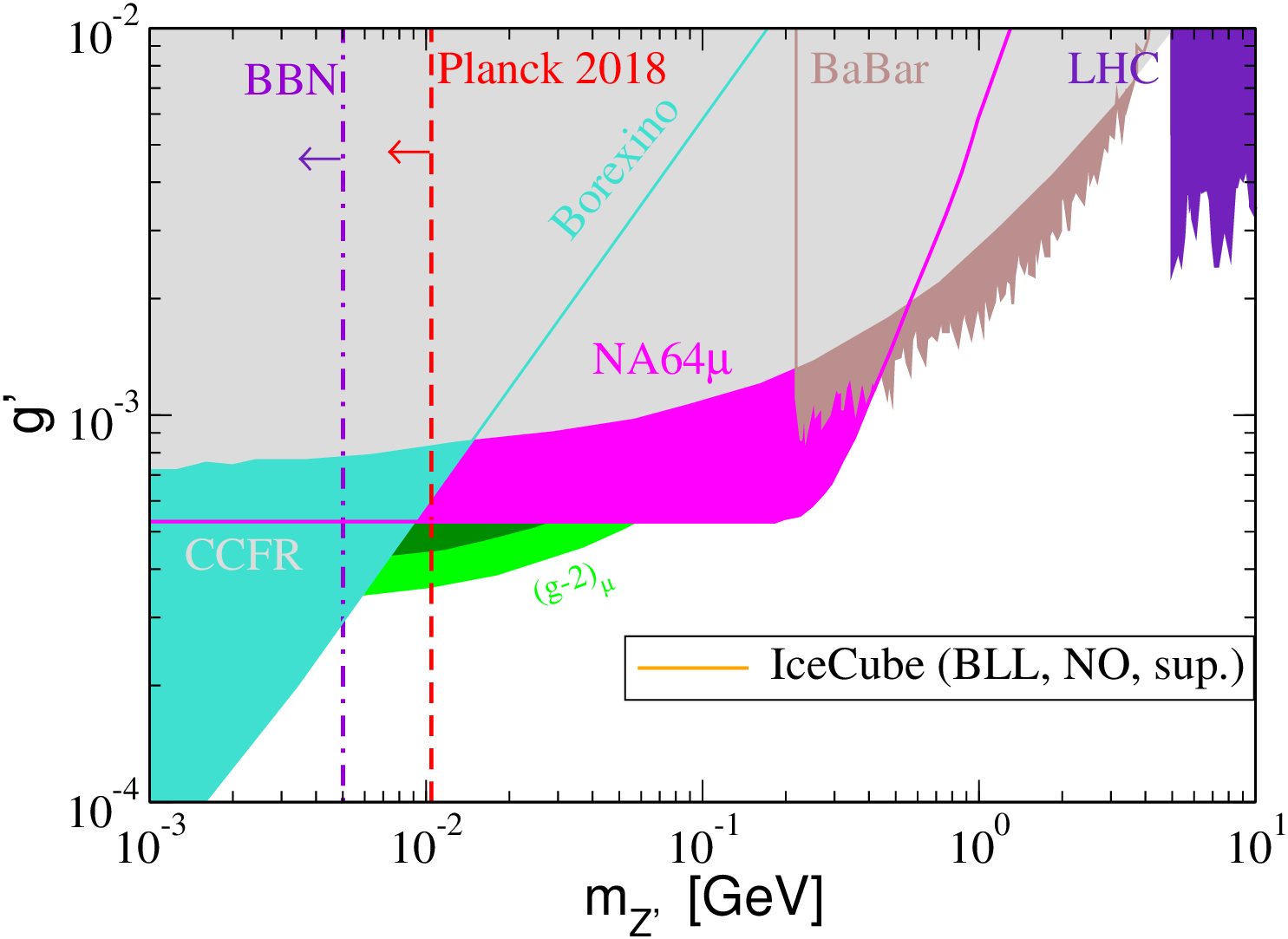}
	\includegraphics[width=0.5\textwidth]{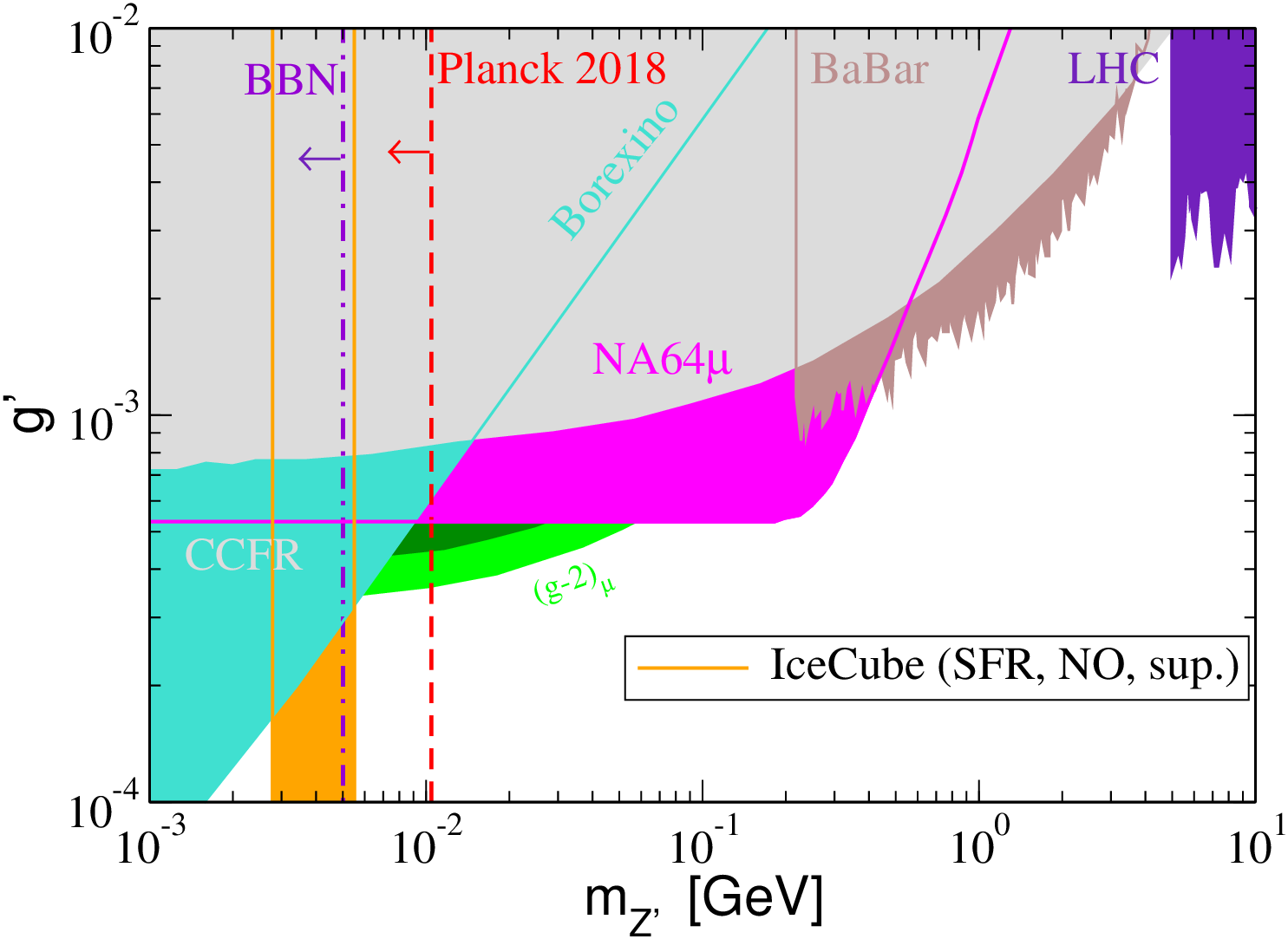} \\
    \includegraphics[width=0.5\textwidth]{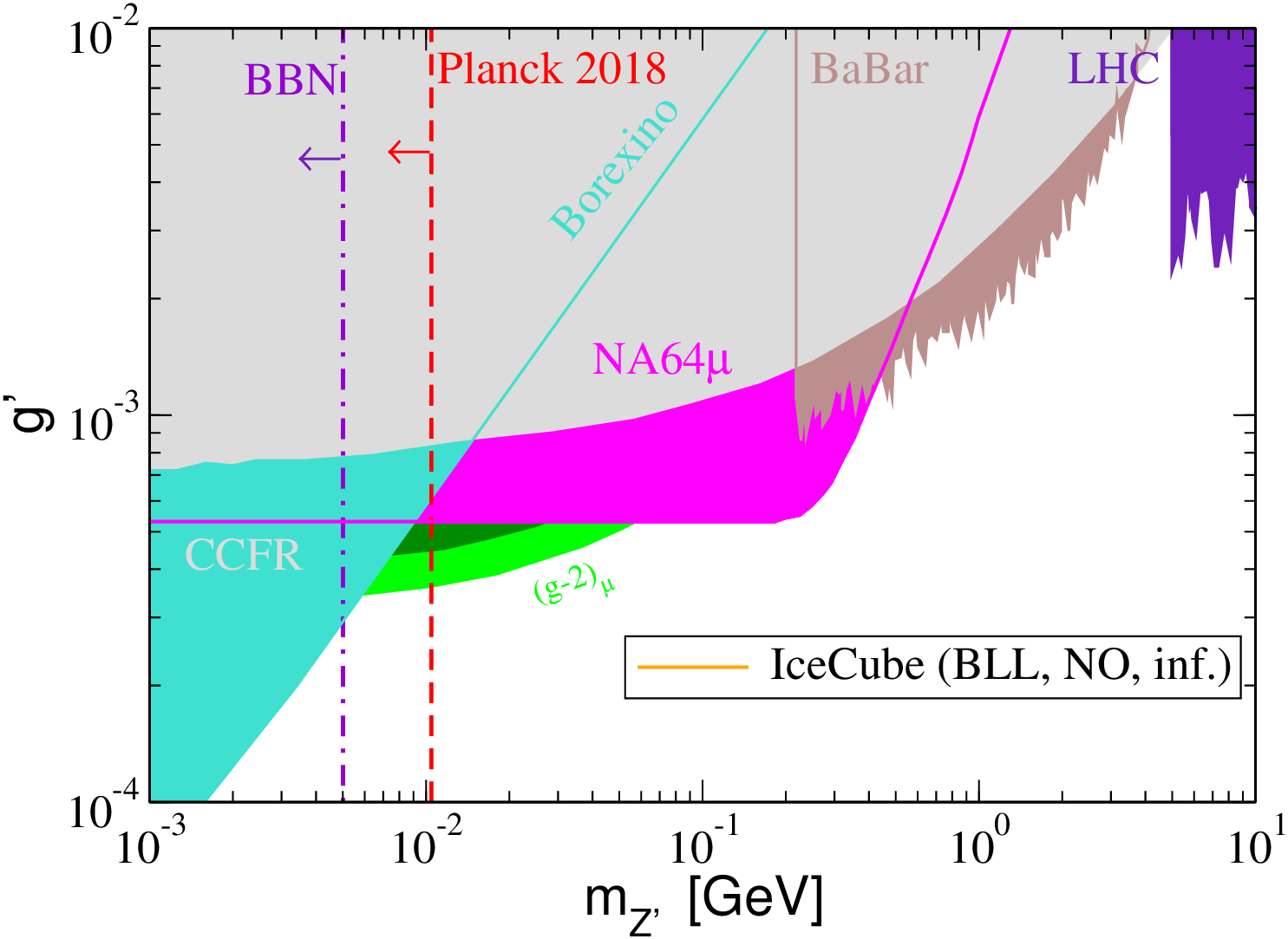}
	\includegraphics[width=0.5\textwidth]{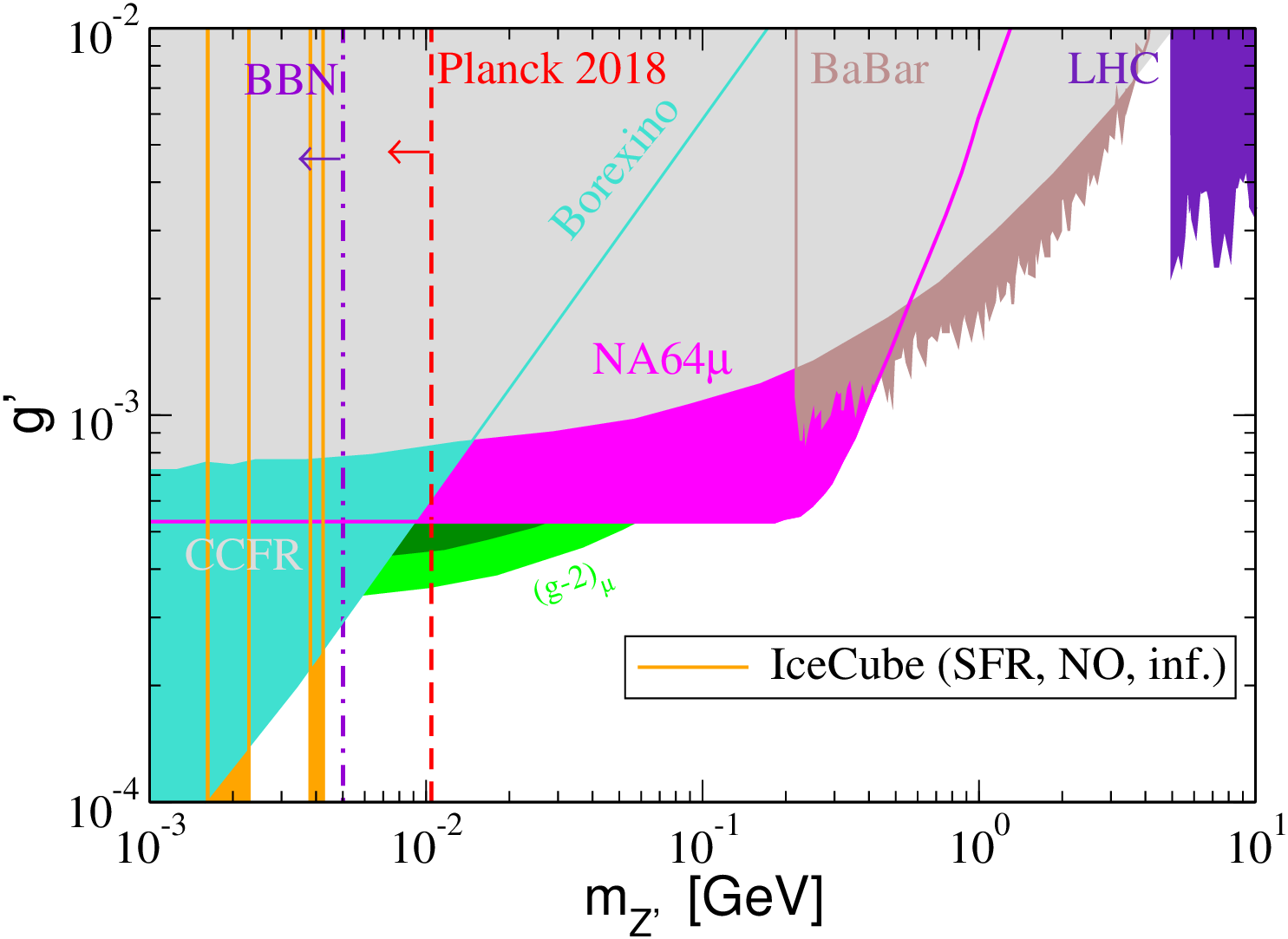}
    
			\end{tabular}
\caption{ Results for the sensitivity of HESE data at IceCube Observatory for the astrophysical neutrino flux to the presence of the $Z'$ gauge boson predicted by the $L_\mu - L_\tau$ model, derived considering  a normal ordering of neutrino masses. Predictions, at the $2\sigma$ level, obtained considering the  redshift distribution of BLL (left panels) and SFR (right panels), and the superior (upper panels) and inferior (lower panels) limits for sum of neutrino masses. For comparison, the existing constraints from other processes and experiments are also presented. The green bands represent the parameter space in which the presence of a $Z'$ solves the muon magnetic moment anomaly at the $1\sigma$ and $2\sigma$ levels.}
\label{fig:spaceNO}
\end{figure}

\begin{figure}[t]
	\centering
	\begin{tabular}{ccc}
	\includegraphics[width=0.5\textwidth]{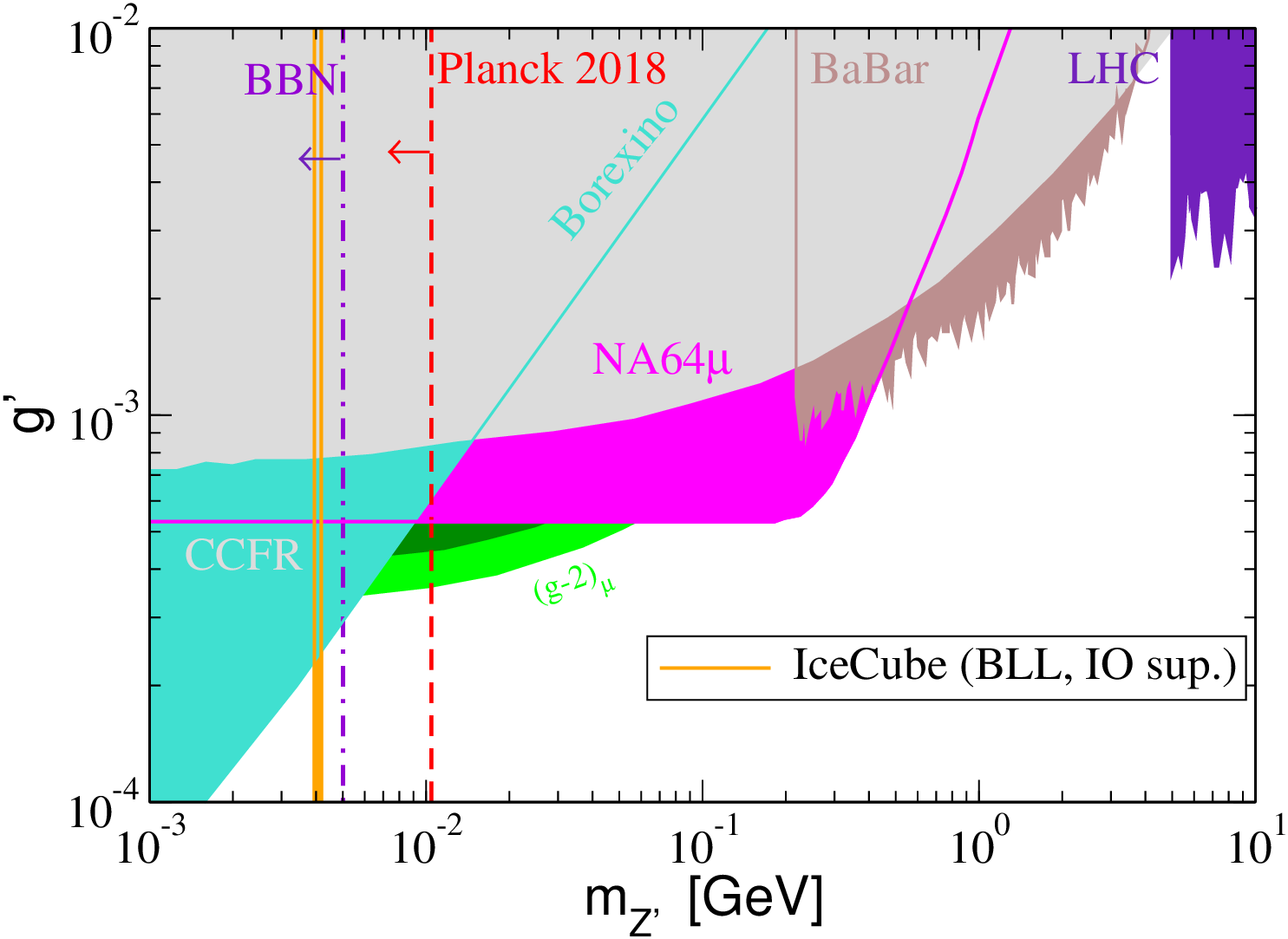}
	\includegraphics[width=0.5\textwidth]{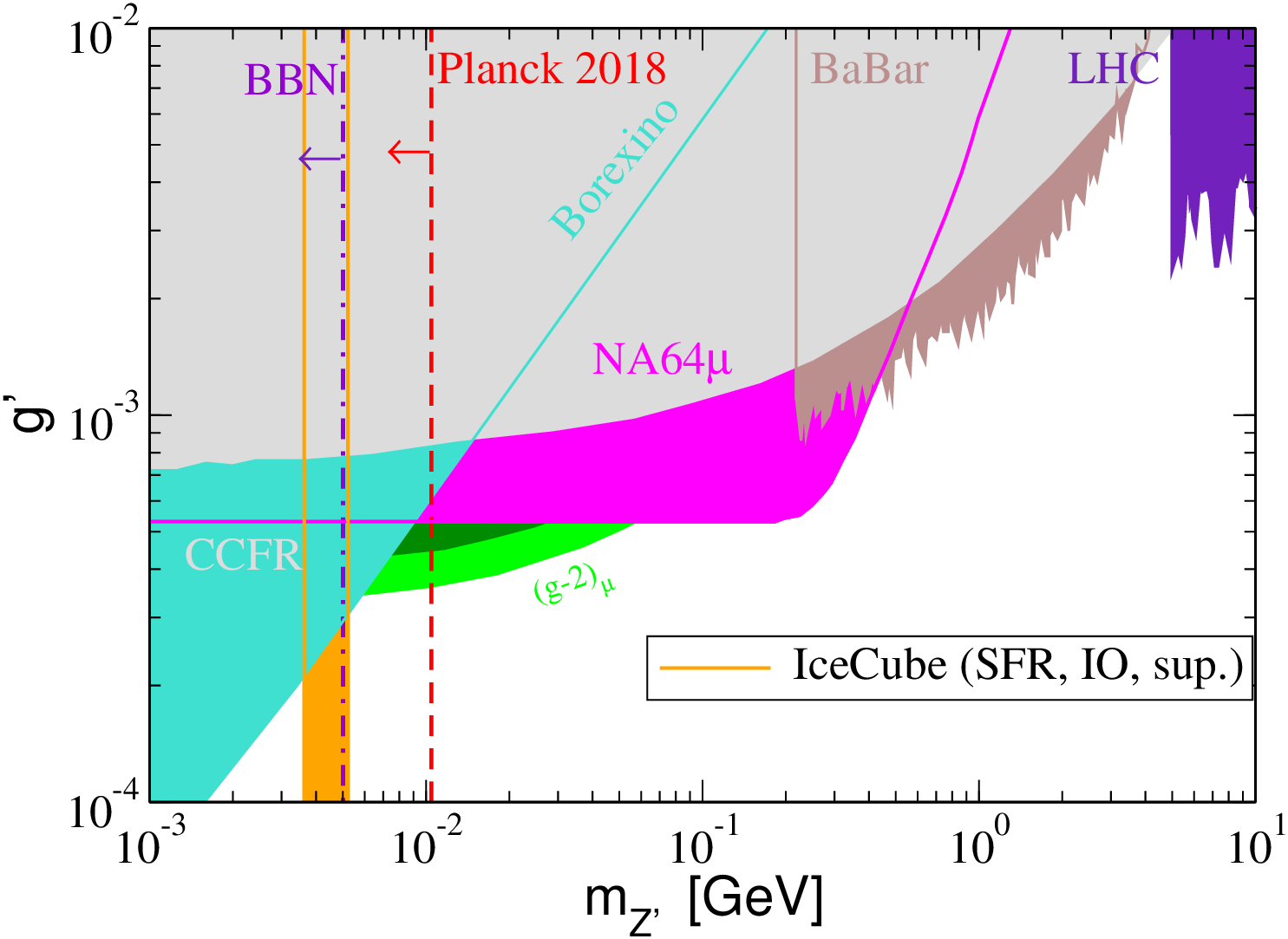} \\
    \includegraphics[width=0.5\textwidth]{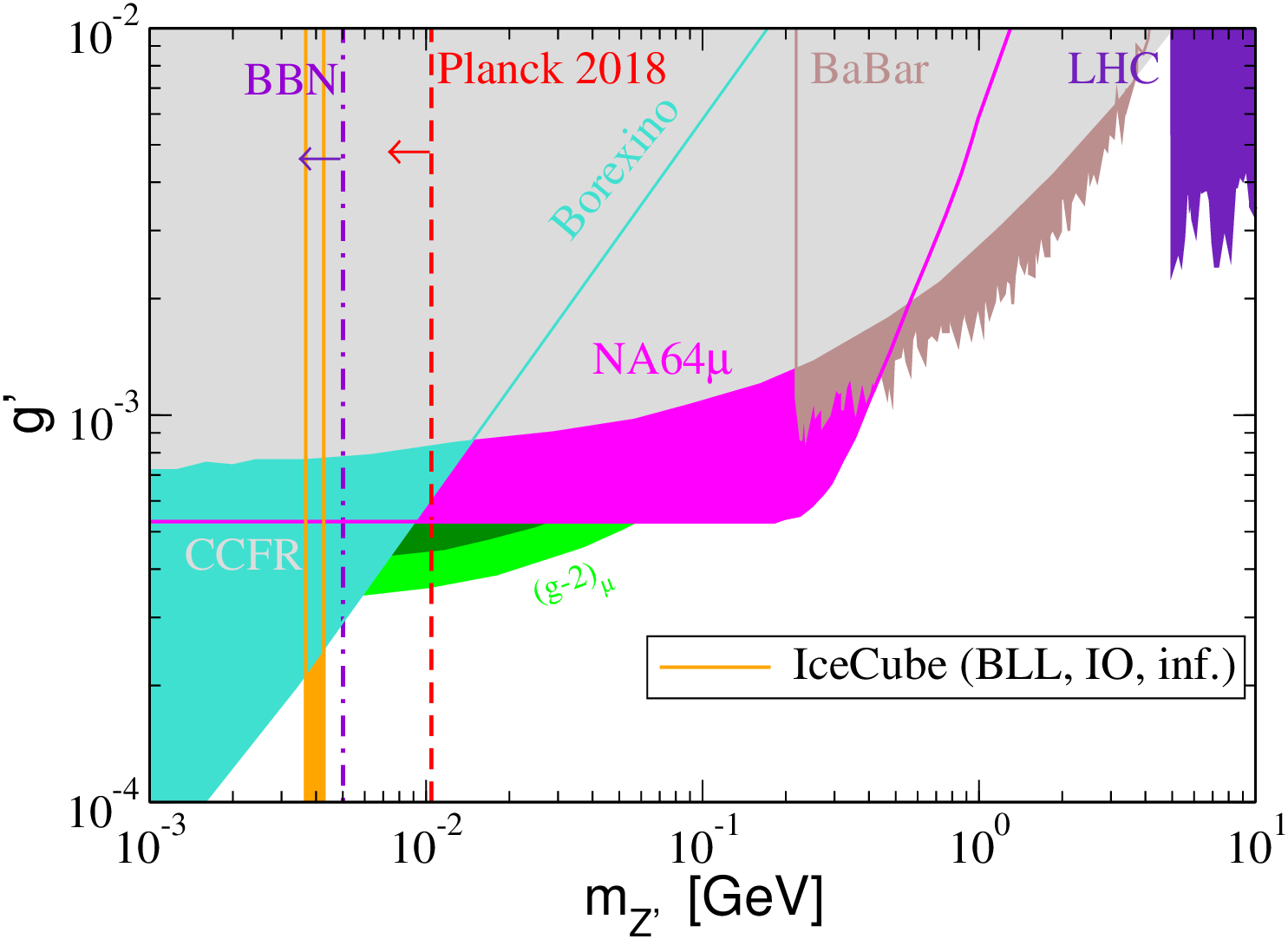}
	\includegraphics[width=0.5\textwidth]{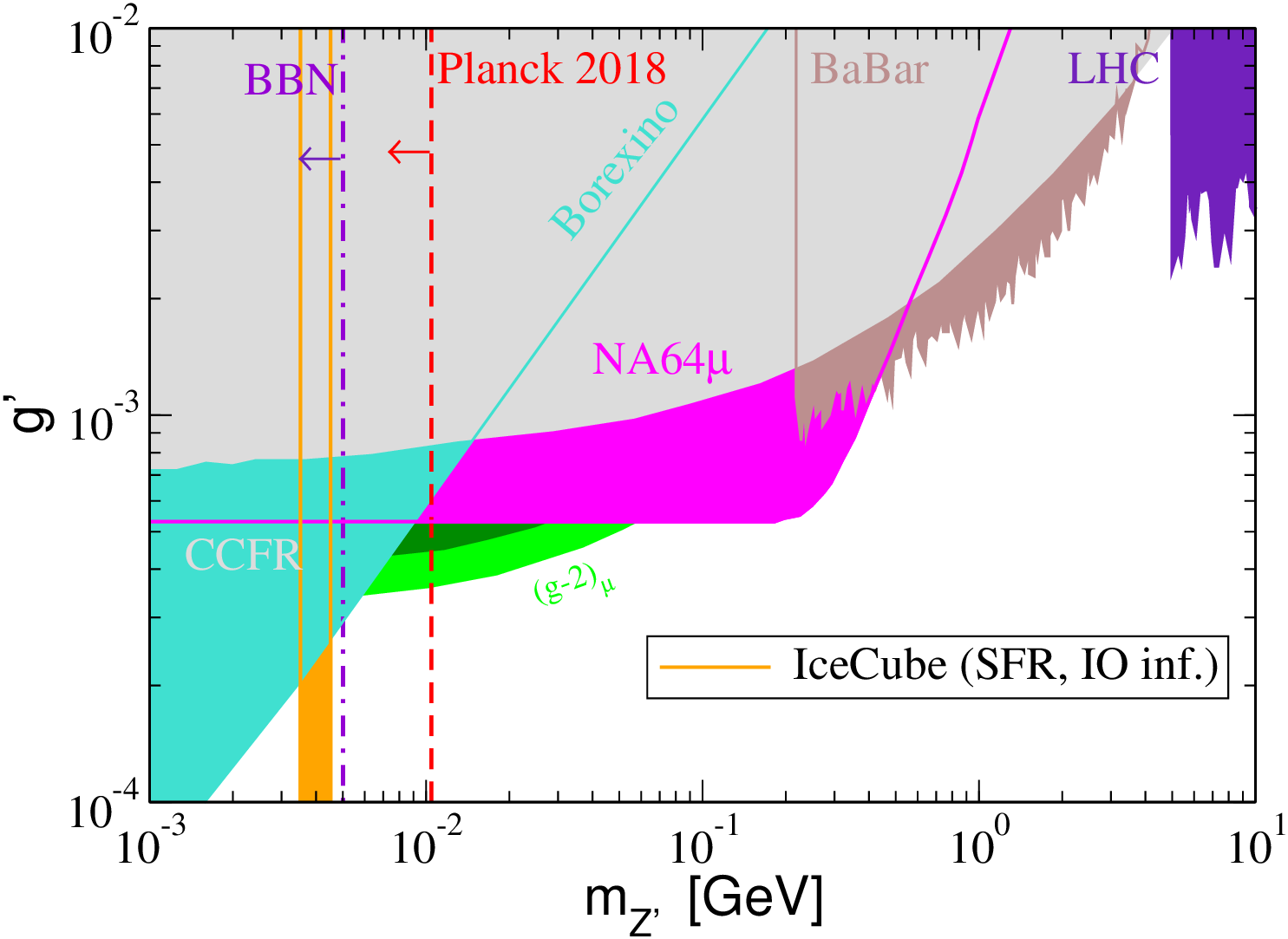}
    
			\end{tabular}
\caption{ Results for the sensitivity of HESE data at IceCube Observatory for the astrophysical neutrino flux to the presence of the $Z'$ gauge boson predicted by the $L_\mu - L_\tau$ model, derived considering  an inverted ordering of neutrino masses. Predictions, at the $2\sigma$ level, obtained considering the  redshift distribution of BLL (left panels) and SFR (right panels), and the superior (upper panels) and inferior (lower panels) limits for sum of neutrino masses. For comparison, the existing constraints from other processes and experiments are also presented. The green bands represent the parameter space in which the presence of a $Z'$ solves the muon magnetic moment anomaly at the $1\sigma$ and $2\sigma$ levels. }
\label{fig:spaceIO}
\end{figure}

In Fig. \ref{fig:spaceNO} we present our results for the sensitivity of the IceCube Observatory, derived considering the HESE data and a normal ordering. Our goal is to derive the corresponding constraints on the mass and coupling parameter space of the $Z'$ boson with a 95\% confidence level. The results obtained  assuming the BLL (SFR) distribution are presented in the left (right) panels. Finally, the upper (lower) panels are for the superior (inferior) limit of the sum of masses. For comparison, the existing bounds derived using other experiments \cite{CMS:2018yxg,ATLAS:2023vxg,ATLAS:2024uvu,BaBar:2016sci,Bellini:2011rx,Harnik:2012ni,NA64:2024klw,Escudero:2019gzq,Ghosh:2024cxi,CCFR:1991lpl} are also presented, as well as the region not yet excluded that explains $(g-2)_\mu$ (green bands) \cite{Keshavarzi:2018mgv} is also shown. One has that for a BLL redshift distribution, the IceCube is not able to probe any new region of the parameter space. In contrast, for a SFR distribution, a small region of coupling and masses, not covered by the Borexino experiment, starts to be probed by IceCube. The region covered is dependent on the ordering assumed, as demonstrated in Fig. \ref{fig:spaceIO}, where we presented the results derived assuming an inverted ordering. In this case, IceCube is also able to probe a $Z^{\prime}$ gauge boson for a BLL distribution. 
 One has that the regions probed by IceCube are, in principle, excluded by cosmological constraints, but it is important to note that these constraints came from indirect measurements and depend on the assumptions present in the standard $\Lambda$CDM cosmology. As a consequence, the possibility of probing this region of the parameter space using a distinct process is timely. 
Another aspect that is important to emphasize is that in the analyzed region of the parameter space, the results are  almost independent of the coupling constant $g'$. This result is implied by the maximum of the cross-section $\nu_i \bar{\nu}_j$ as a function of the energy of the incident neutrino (see Eq. \ref{eq:cs}) being independent of $g'$.
 However, as the width of the cross-section distribution is proportional to $g'^{2}$,  the distribution becomes narrower as $g'$ decreases and the impact of a $Z^{\prime}$ vanishes when $g' \rightarrow 0$. Our results were obtained for $g' > 10^{-4}$, but recent analyses shown that in the region of $g'< 10^{-4}$ the cross-section becomes so narrow that the impact on the attenuation of the neutrino flux become negligible (see Refs. \cite{Carpio:2021jhu,Hooper:2023fqn,delaVega:2024pbk}).

\begin{figure}[t]
	\centering
	\begin{tabular}{ccc}
	\includegraphics[width=0.5\textwidth]{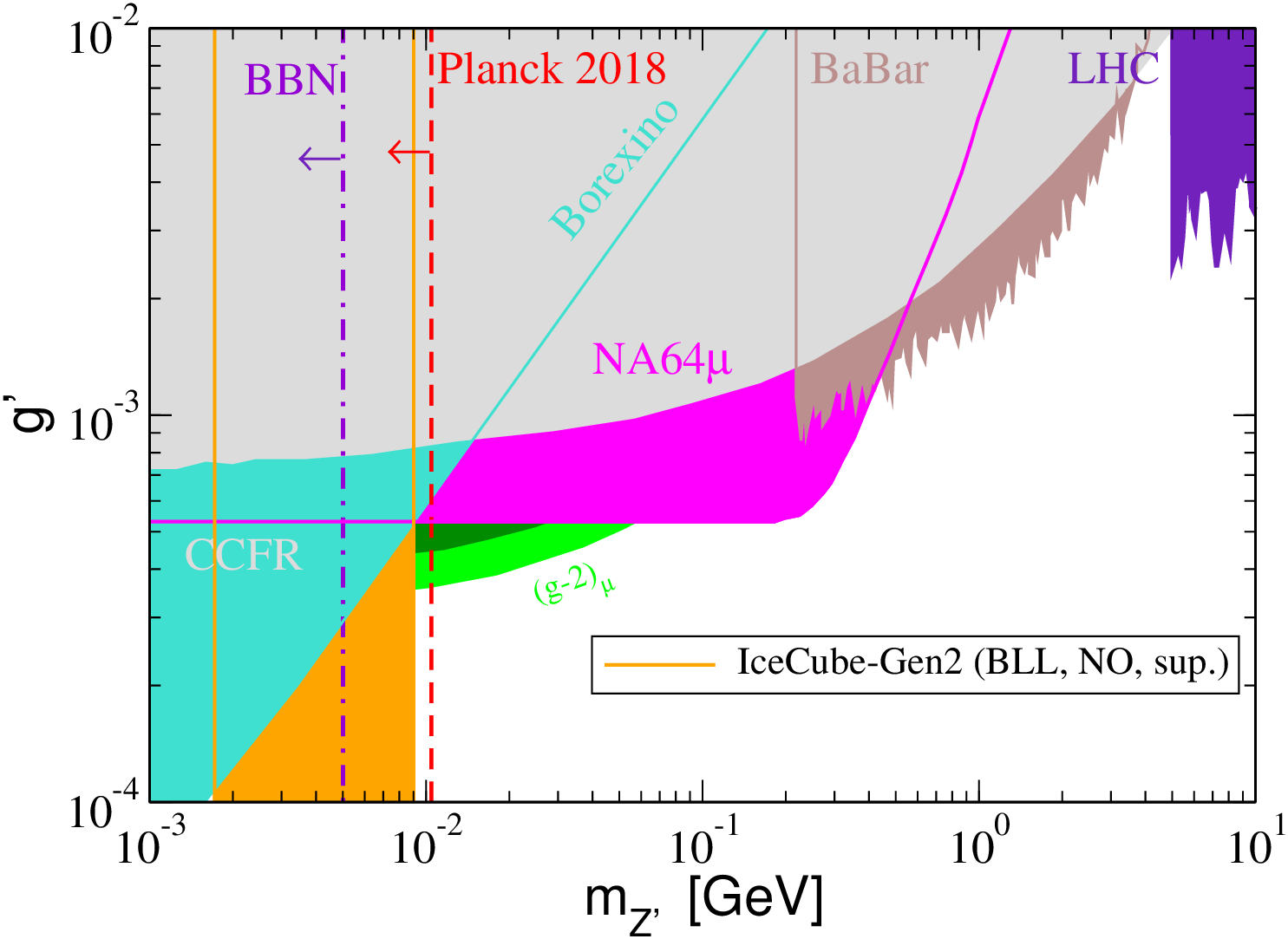}
	\includegraphics[width=0.5\textwidth]{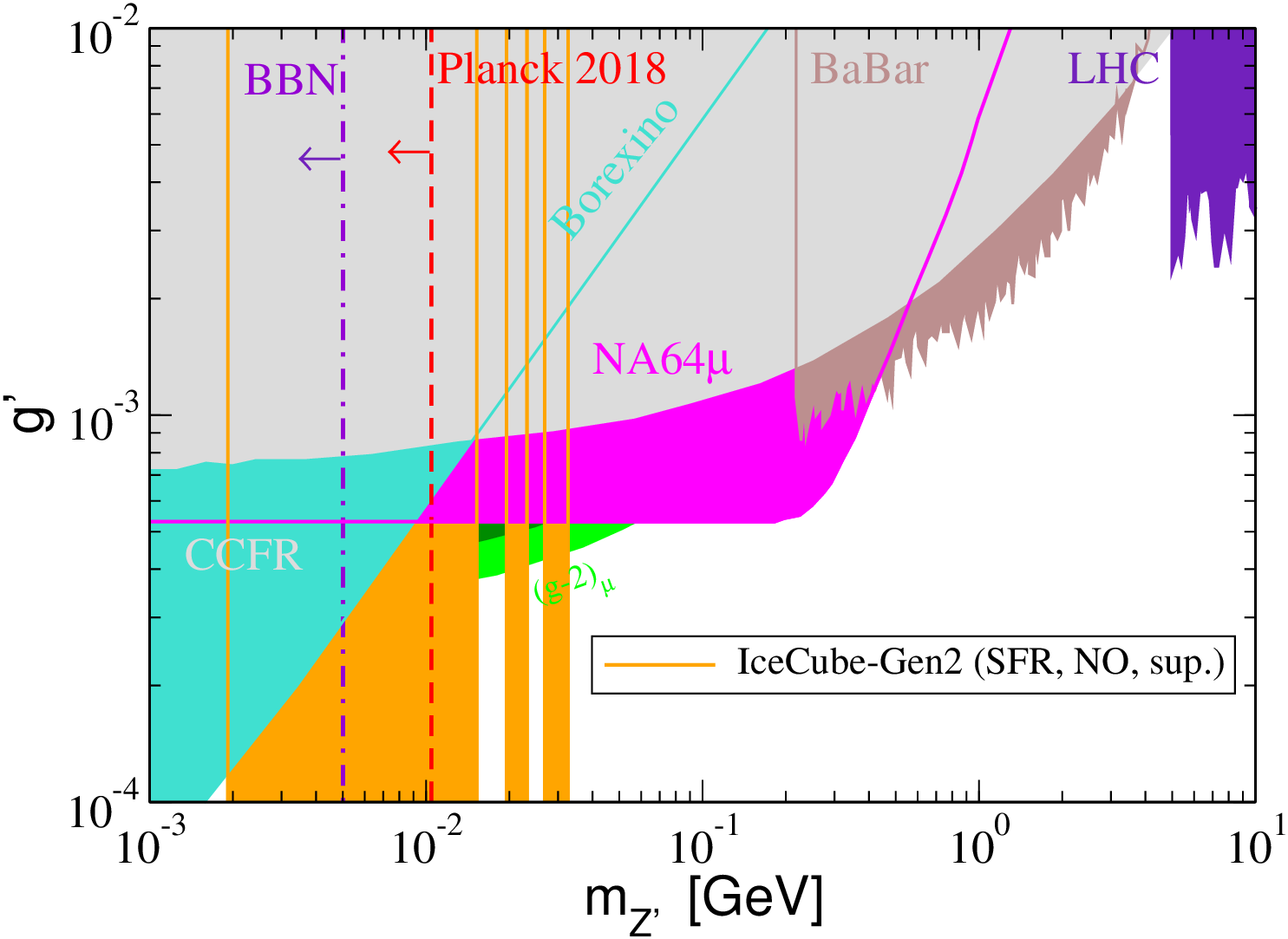} \\
    \includegraphics[width=0.5\textwidth]{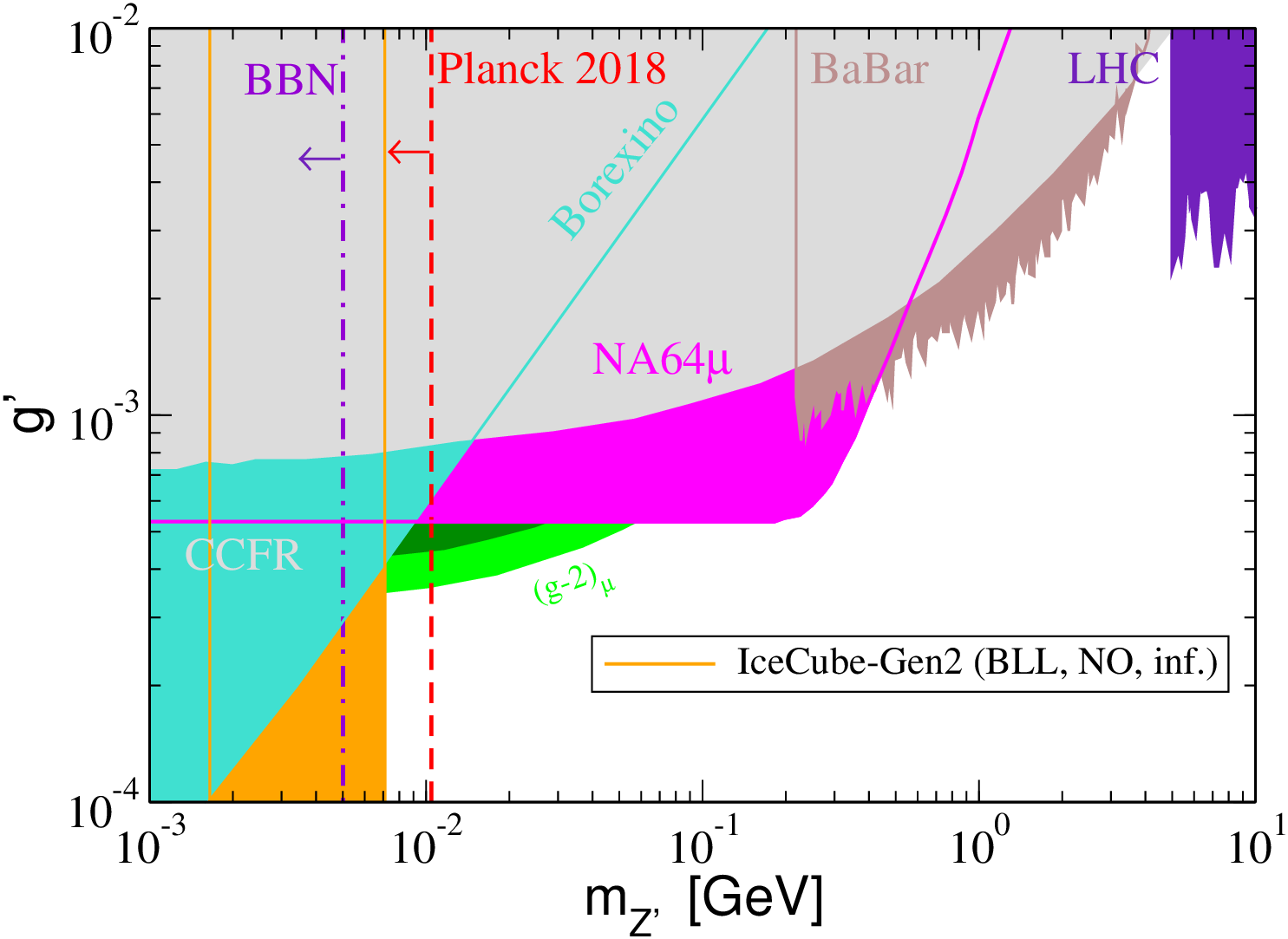}
	\includegraphics[width=0.5\textwidth]{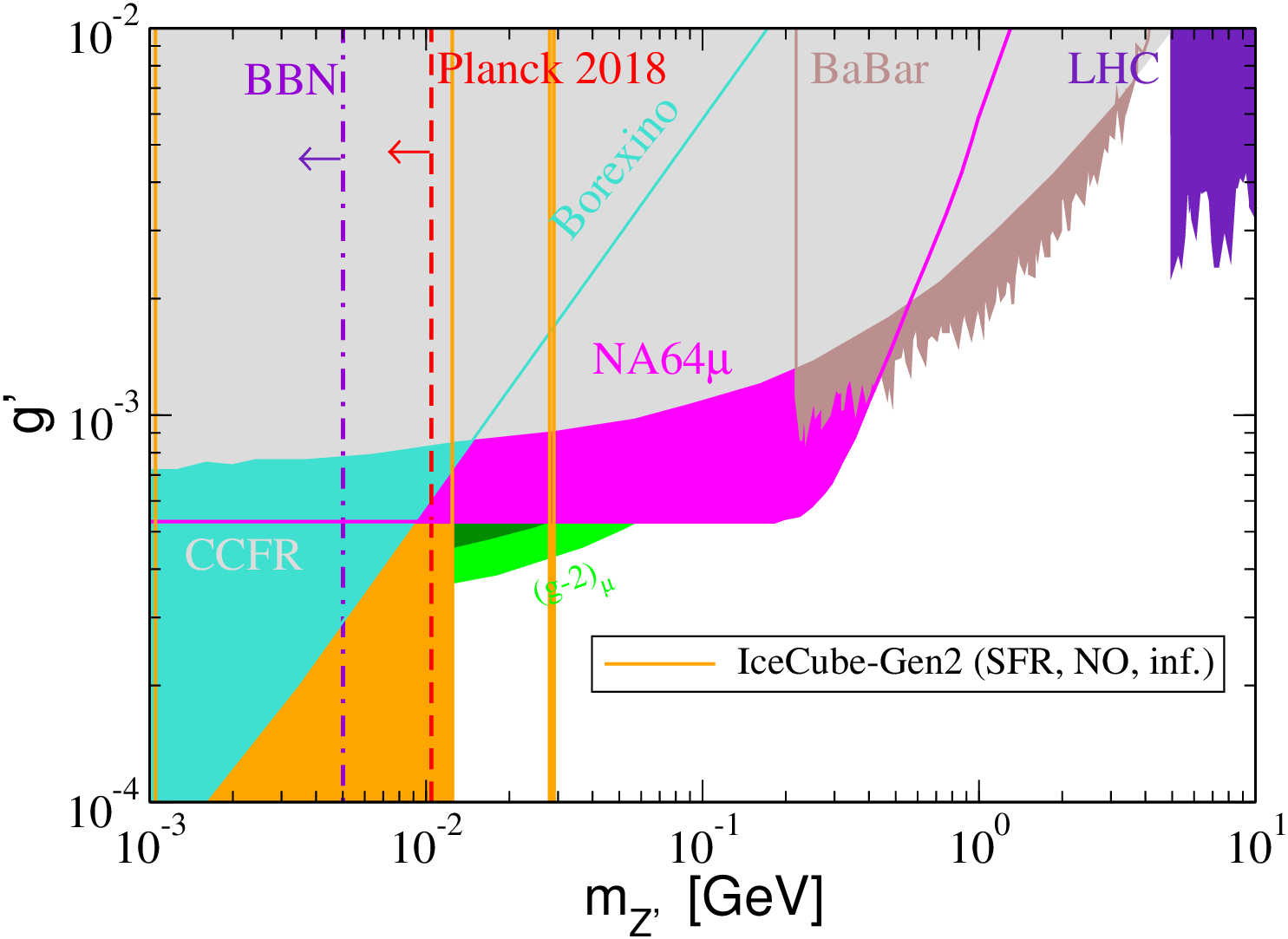}
    
			\end{tabular}
\caption{Results for the sensitivity of HESE data expected at  IceCube-Gen2 Observatory for the astrophysical neutrino flux to the presence of the $Z'$ gauge boson predicted by the $L_\mu - L_\tau$ model, derived considering  a  normal ordering of neutrino masses. Predictions, at the $2\sigma$ level, obtained considering the  redshift distribution of BLL (left panels) and SFR (right panels), and the superior (upper panels) and inferior (lower panels) limits for sum of neutrino masses. For comparison, the existing constraints from other processes and experiments are also presented. The green bands represent the parameter space in which the presence of a $Z'$ solves the muon magnetic moment anomaly at the $1\sigma$ and $2\sigma$ levels. }
\label{fig:spaceNOgen2}
\end{figure}

\begin{figure}[t]
	\centering
	\begin{tabular}{ccc}
	\includegraphics[width=0.5\textwidth]{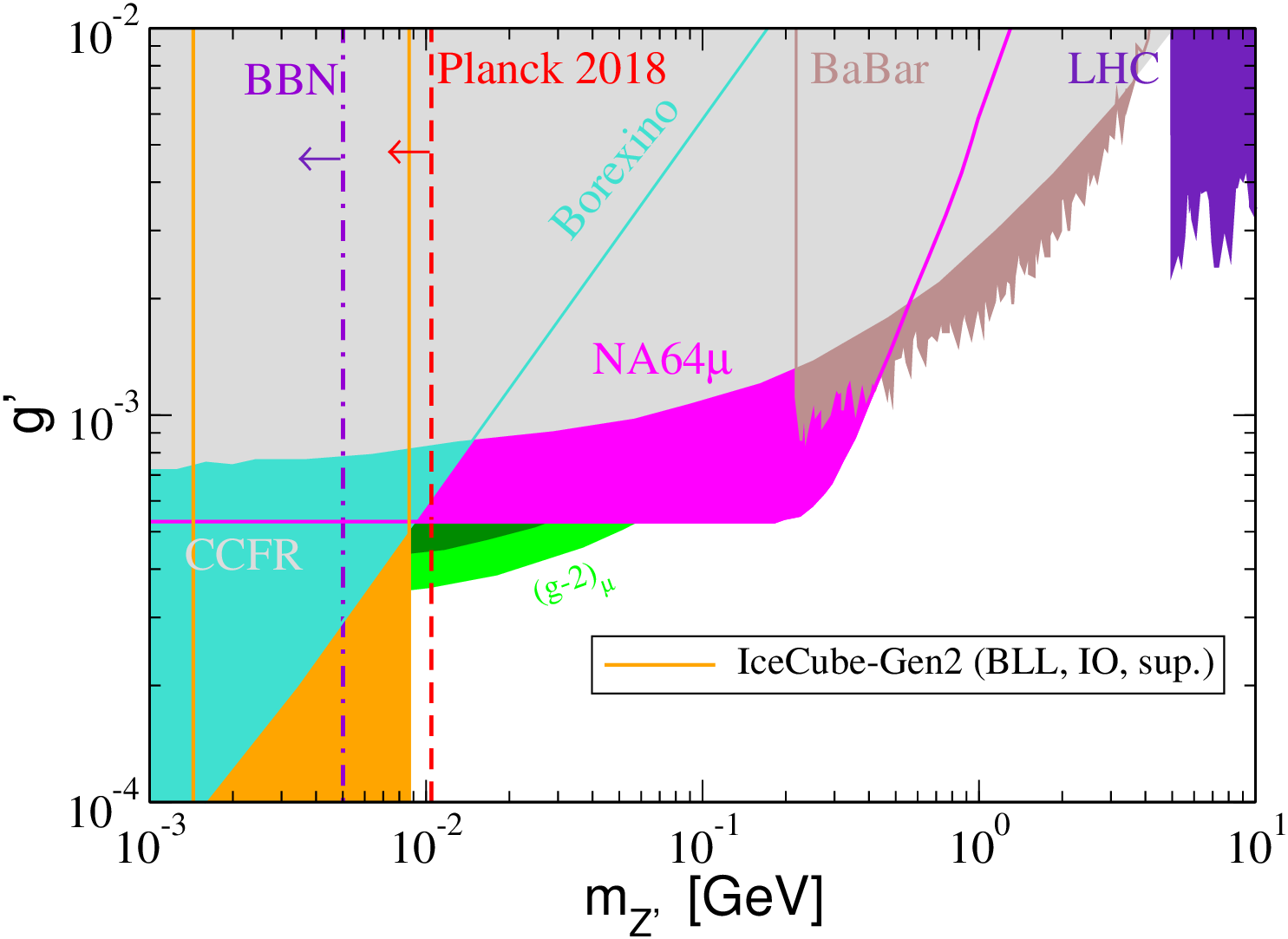}
	\includegraphics[width=0.5\textwidth]{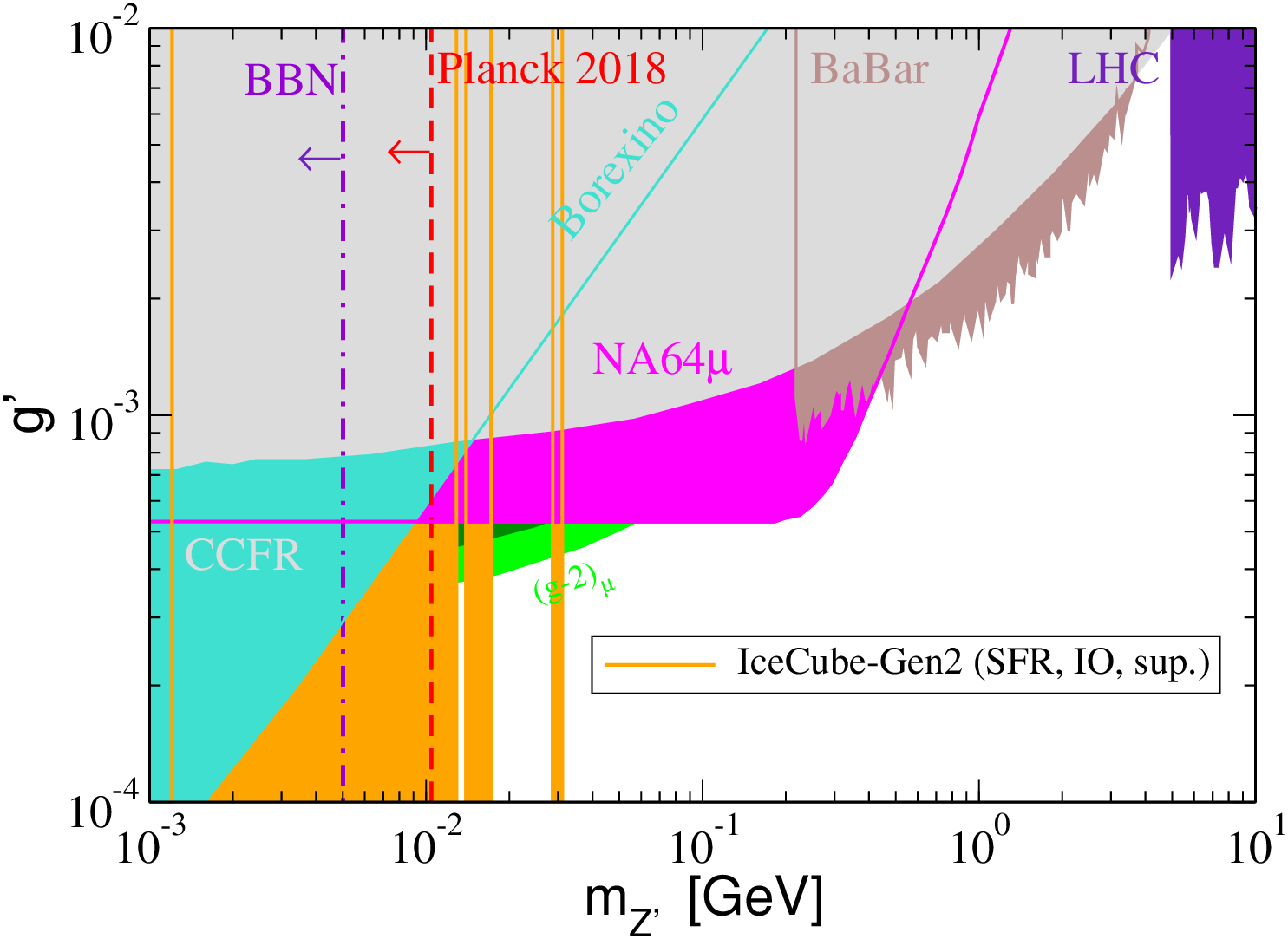} \\
    \includegraphics[width=0.5\textwidth]{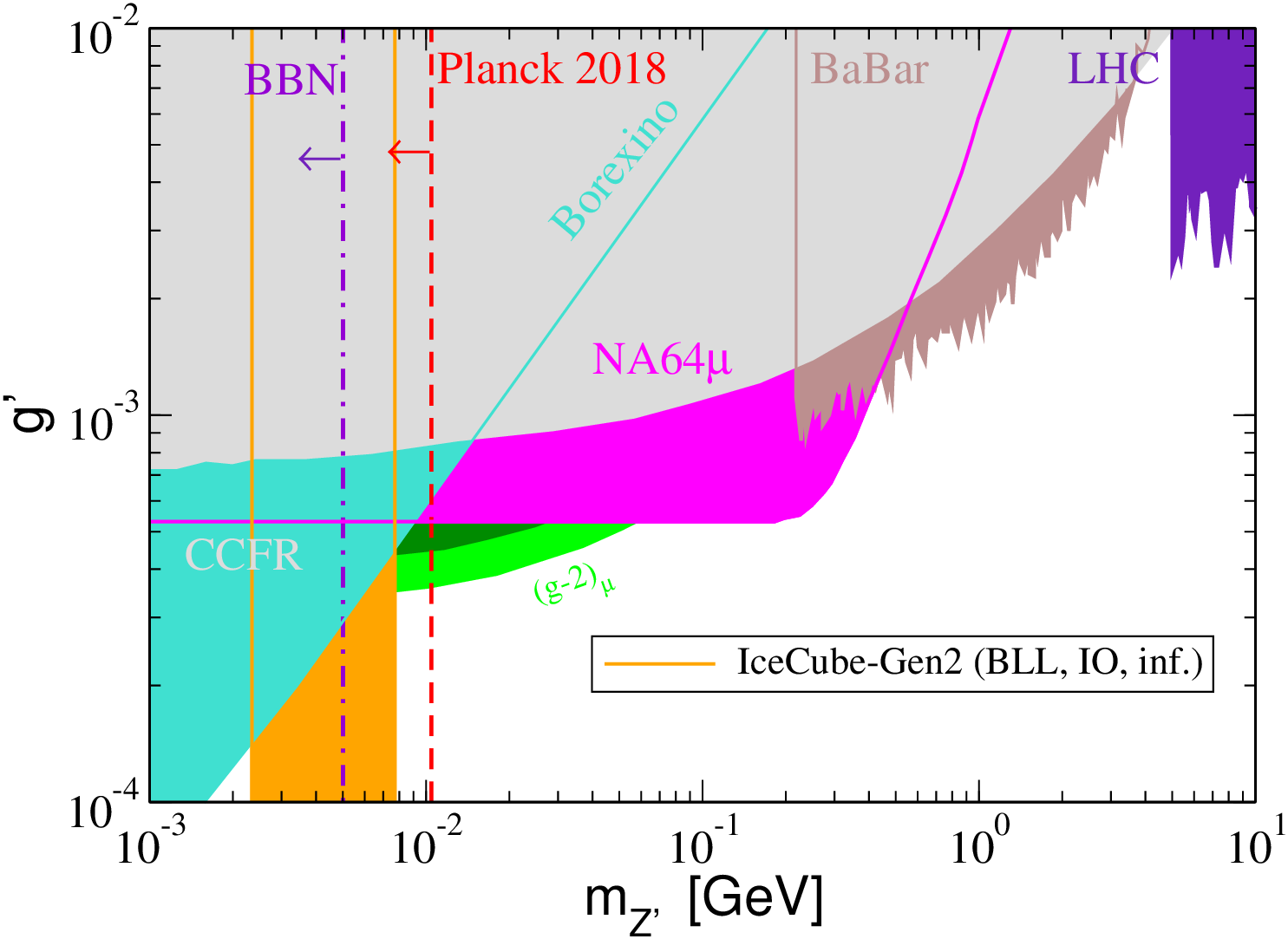}
	\includegraphics[width=0.5\textwidth]{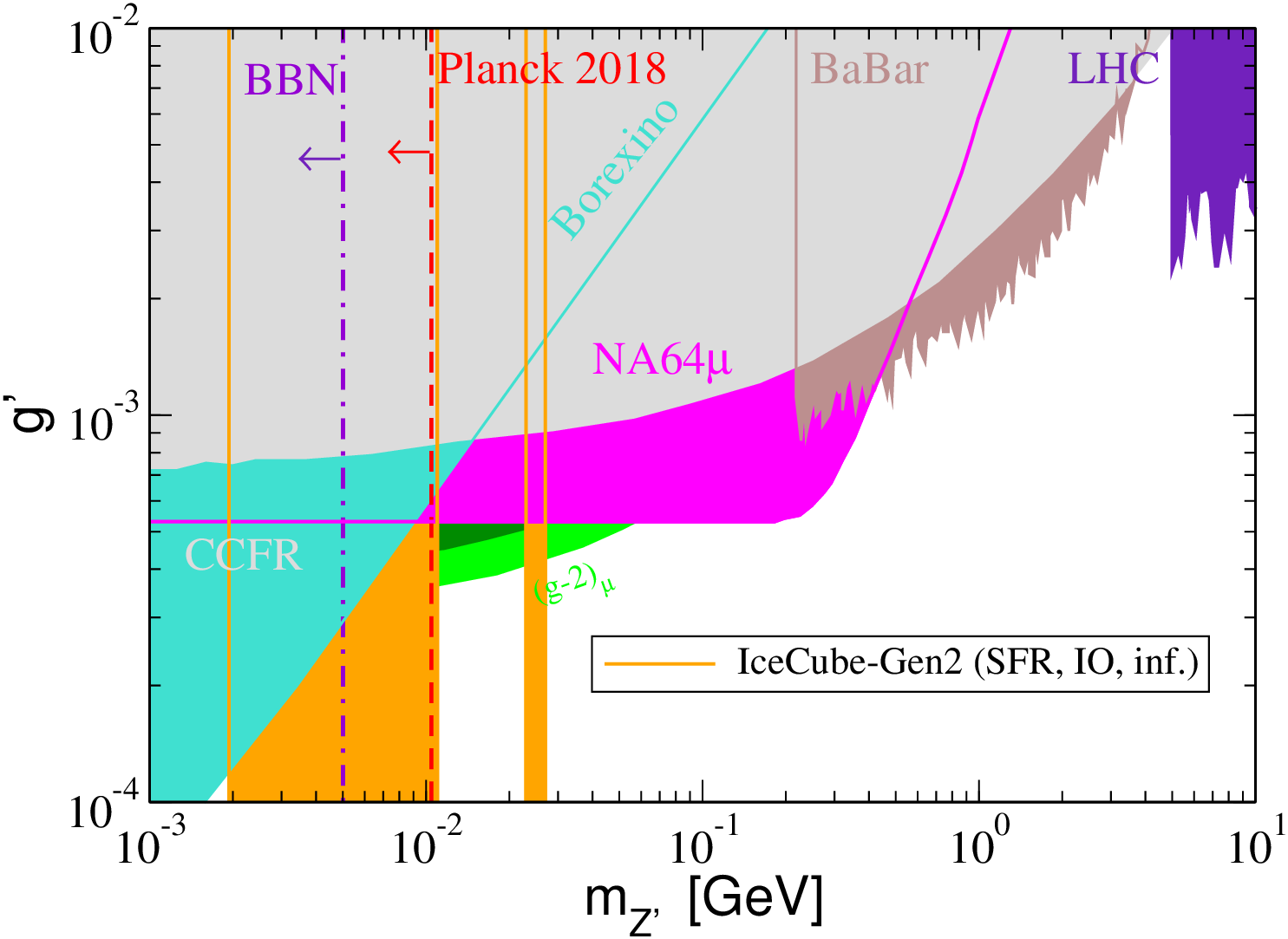}
    
			\end{tabular}
\caption{ Results for the sensitivity of HESE data expected at  IceCube-Gen2 Observatory for the astrophysical neutrino flux to the presence of the $Z'$ gauge boson predicted by the $L_\mu - L_\tau$ model, derived considering  an inverted ordering of neutrino masses. Predictions, at the $2\sigma$ level, obtained considering the  redshift distribution of BLL (left panels) and SFR (right panels), and the superior (upper panels) and inferior (lower panels) limits for sum of neutrino masses. For comparison, the existing constraints from other processes and experiments are also presented. The green bands represent the parameter space in which the presence of a $Z'$ solves the muon magnetic moment anomaly at the $1\sigma$ and $2\sigma$ levels.  }
\label{fig:spaceIOgen2}
\end{figure}

Finally, let's extend our analysis for the IceCube-Gen2, which is the planned upgrade of  the IceCube observatory that will  amplify the number of observed neutrino events. In our analysis, we will consider the same time of exposure currently at the IceCube and an increasing in its volume by a factor 8. Furthermore, we will consider the same parameters for systematic uncertainties in the Pull method used in IceCube. The corresponding results for the sensitivity of the IceCube-Gen2 Observatory are presented in Figs. \ref{fig:spaceNOgen2} and \ref{fig:spaceIOgen2}, for the normal and inverted ordering, respectively.
Our results indicate that IceCube-Gen2 will cover a region in the parameter space not excluded by the Borexino and NA64$\mu$ experiments, independent of the ordering considered. Moreover, it will cover part of the region  between 10 MeV and approximately 30 MeV for the $Z'$ mass, including part of the region that can explain $(g- 2)_\mu$, which currently remains not excluded. 
 In the case of the results derived considering the SFR distribution, the region covered by the IceCube-Gen2 is not continuous. This occurs because the sensitivity to the Glashow resonance covers an energy range that would not be excluded if only charged current interactions were taken into account. Our results further indicate that the excluded regions in IceCube-Gen2 will  weakly depend on the mass ordering, but will be larger for the superior limit case.

\section{Summary}
\label{sec:sum}

In this work, we have studied the impact of a $Z^{\prime}$ gauge boson, predicted by the $L_\mu - L_\tau$ model, on the propagation of astrophysical neutrinos until they reach neutrino observatories as the IceCube. We investigated how the existence of this new boson impacts the description of HESE data in IceCube
considering different assumption for the redshift distributions of the neutrino sources, distinct ordering for the mass eigenstates and different limits for the sum of neutrino masses. We have obtained the current sensitivity of IceCube and estimated the sensitivity for the IceCube-Gen2 to the new physics considering different values of mass and coupling for the $Z'$ gauge boson.  Our results indicated that the description of the HESE data is subtly improved when a low - mass $Z^{\prime}$ is present and that IceCube is able to probe the region of small coupling and masses of the order of few MeV. Moreover,  we demonstrated that  IceCube-Gen2 can increase the excluded region of mass and coupling of the $Z'$ boson, not previously covered by the Borexino and NA64$\mu$ experiments.

\begin{acknowledgments}
 R. F. acknowledges support from the Conselho Nacional de Desenvolvimento Cient\'{\i}fico e Tecnol\'ogico (CNPq, Brazil), Grant No. 161770/2022-3. V.P.G. was partially supported by CNPq, FAPERGS and INCT-FNA (Process No. 464898/2014-5). D.R.G. was partially supported by CNPq and MCTI.

\end{acknowledgments}

\hspace{1.0cm}


\begin{thebibliography}{99}


     \bibitem{HESE1}
M.~G.~Aartsen {\it et al.} [IceCube],
 Science {\bf 342}, 1242856 (2013)

     \bibitem{HESE2}
R.~Abbasi {\it et al.} [IceCube],
 Phys.\ Rev.\ D {\bf 104}, 022002 (2021)

\bibitem{IceCube:2024fxo}
R.~Abbasi \textit{et al.} [IceCube],
Phys. Rev. D \textbf{110}, no.2, 022001 (2024)

       \bibitem{tracksNorte}
P.~Fuerst {\it et al.} [IceCube],
 Pos {\bf ICRC2023} 1046 (2023)

     \bibitem{cascades}
M.~G.~Aartsen {\it et al.} [IceCube],
 Phys.\ Rev.\ Lett. {\bf 125}, 121104 (2020)


 \bibitem{blazar1}
M.~G.~Aartsen {\it et al.} [IceCube], 
Science 361, 6398 (2018);

 \bibitem{blazar2}
M.~G.~Aartsen {\it et al.} [IceCube, Fermi-LAT, MAGIC, AGILE, ASAS-SN, HAWC, H.E.S.S., INTEGRAL, Kanata, Kiso,
Kapteyn, Liverpool Telescope, Subaru, Swift NuSTAR, VERITAS, VLA/17B-403 Collaboration], 
Science 361, 6398 (2018);

  	
 \bibitem{sn1068}
R.~Abbasi {\it et al.} [IceCube], 
Science 378, 6619 (2022);

\bibitem{IceCube-Gen2:2020qha}
M.~G.~Aartsen \textit{et al.} [IceCube-Gen2],
J. Phys. G \textbf{48}, no.6, 060501 (2021)

\bibitem{Goncalves:2022uwy}
V.~P.~Goncalves, D.~R.~Gratieri and A.~S.~C.~Quadros,
Eur. Phys. J. C \textbf{82}, no.11, 1011 (2022)


\bibitem{IceCube:2018tkk}
M.~G.~Aartsen \textit{et al.} [IceCube],
Eur. Phys. J. C \textbf{78}, no.10, 831 (2018)

\bibitem{Berlin:2024lwe}
A.~Berlin and D.~Hooper,
Phys. Rev. D \textbf{110}, no.7, 075018 (2024)

\bibitem{Shoemaker:2015qul}
I.~M.~Shoemaker and K.~Murase,
Phys. Rev. D \textbf{93}, no.8, 085004 (2016)

\bibitem{Bustamante:2020mep}
M.~Bustamante, C.~Rosenstr\o{}m, S.~Shalgar and I.~Tamborra,
Phys. Rev. D \textbf{101}, no.12, 123024 (2020)

\bibitem{Agarwalla:2023sng}
S.~K.~Agarwalla, M.~Bustamante, S.~Das and A.~Narang,
JHEP \textbf{08}, 113 (2023)

\bibitem{KA:2023dyz}
S.~K.A., A.~Das, G.~Lambiase, T.~Nomura and Y.~Orikasa,
Eur. Phys. J. C \textbf{84}, no.11, 1224 (2024)

\bibitem{Wang:2025qap}
I.~R.~Wang, X.~J.~Xu and B.~Zhou,
[arXiv:2501.07624 [hep-ph]].

\bibitem{Esteban:2021tub}
I.~Esteban, S.~Pandey, V.~Brdar and J.~F.~Beacom,
Phys. Rev. D \textbf{104}, no.12, 123014 (2021)

\bibitem{He:1990pn}
X.~G.~He, G.~C.~Joshi, H.~Lew and R.~R.~Volkas,
Phys. Rev. D \textbf{43} (1991), 22-24

\bibitem{He:1991qd}
X.~G.~He, G.~C.~Joshi, H.~Lew and R.~R.~Volkas,
Phys. Rev. D \textbf{44} (1991), 2118-2132

\bibitem{Keshavarzi:2018mgv}
A.~Keshavarzi, D.~Nomura and T.~Teubner,
Phys. Rev. D \textbf{97}, no.11, 114025 (2018)

\bibitem{Kamada:2015era}
A.~Kamada and H.~B.~Yu,
Phys. Rev. D \textbf{92}, no.11, 113004 (2015)

\bibitem{Araki:2015mya}
T.~Araki, F.~Kaneko, T.~Ota, J.~Sato and T.~Shimomura,
Phys. Rev. D \textbf{93}, no.1, 013014 (2016)

\bibitem{DiFranzo:2015qea}
A.~DiFranzo and D.~Hooper,
Phys. Rev. D \textbf{92}, no.9, 095007 (2015)

\bibitem{Carpio:2021jhu}
J.~A.~Carpio, K.~Murase, I.~M.~Shoemaker and Z.~Tabrizi,
Phys. Rev. D \textbf{107}, no.10, 103057 (2023)

\bibitem{Hooper:2023fqn}
D.~Hooper, J.~Iguaz Juan and P.~D.~Serpico,
Phys. Rev. D \textbf{108}, no.2, 023007 (2023)

\bibitem{delaVega:2024pbk}
L.~M.~G.~de la Vega, E.~Peinado and J.~Wudka,
Phys. Rev. D \textbf{110}, no.9, 095032 (2024)



\bibitem{Bhattacharjee:1999mup}
P.~Bhattacharjee and G.~Sigl,
Phys. Rept. \textbf{327}, 109-247 (2000)


\bibitem{IceCube:2023sov}
R.~Abbasi \textit{et al.} [IceCube],
PoS \textbf{ICRC2023}, 1030 (2023)

\bibitem{Planck:2018vyg}
N.~Aghanim \textit{et al.} [Planck],
Astron. Astrophys. \textbf{641}, A6 (2020)
[erratum: Astron. Astrophys. \textbf{652}, C4 (2021)]

\bibitem{Yuksel:2008cu}
H.~Yuksel, M.~D.~Kistler, J.~F.~Beacom and A.~M.~Hopkins,
Astrophys. J. Lett. \textbf{683}, L5-L8 (2008)

\bibitem{Ajello:2013lka}
M.~Ajello, R.~W.~Romani, D.~Gasparrini, M.~S.~Shaw, J.~Bolmer, G.~Cotter, J.~Finke, J.~Greiner, S.~E.~Healey and O.~King, \textit{et al.}
Astrophys. J. \textbf{780}, 73 (2014)

\bibitem{Esteban:2024eli}
I.~Esteban, M.~C.~Gonzalez-Garcia, M.~Maltoni, I.~Martinez-Soler, J.~P.~Pinheiro and T.~Schwetz,
JHEP \textbf{12}, 216 (2024)


\bibitem{ntargets}
Nancy Wandkowsky for the IceCube Collaboration, TeVPA (2017)

\bibitem{Francener:2024rjw}
R.~Francener, V.~P.~Goncalves and D.~R.~Gratieri,
[arXiv:2403.16611 [hep-ph]].

\bibitem{Reno:2023sdm}
M.~H.~Reno,
Ann. Rev. Nucl. Part. Sci. \textbf{73}, no.1, 181-204 (2023)


 \bibitem{paschos}
E.~A.~Paschos and J.~Y.~Yu,
 Phys.\ Rev.\ D {\bf 65}, 033002 (2002)

 \bibitem{reno}
S.~Kretzer and M.~H.~Reno,
 Phys.\ Rev.\ D {\bf 66}, 113007 (2002)

     
 \bibitem{ct14} S.~Dulat {\it et al.}, Phys. Rev. D {\bf 93}, 033006 (2016).

\bibitem{Glashow:1960zz}
S.~L.~Glashow,
Phys. Rev. \textbf{118}, 316-317 (1960)

\bibitem{Poisson:1837}
S.~D.~Poisson,
Bachelier (1837)

\bibitem{Neyman:1933wgr}
J.~Neyman and E.~S.~Pearson,
Phil. Trans. Roy. Soc. Lond. A \textbf{231}, no.694-706, 289-337 (1933)

\bibitem{Fogli:2002pt}
G.~L.~Fogli, E.~Lisi, A.~Marrone, D.~Montanino and A.~Palazzo,
Phys. Rev. D \textbf{66}, 053010 (2002)

\bibitem{IceCube:2015gsk}
M.~G.~Aartsen \textit{et al.} [IceCube],
Astrophys. J. \textbf{809}, no.1, 98 (2015)

\bibitem{Wilks:1938dza}
S.~S.~Wilks,
Annals Math. Statist. \textbf{9}, no.1, 60-62 (1938)

\bibitem{Akaike}{ Akaike, H., A new look at the statistical model identification. IEEE Transactions on Automatic Control, Boston, v.19, n.6, p.716–723, Dec. 1974.}

\bibitem{Burnham} {Burnham, K.P.,   Anderson, D.R. (2002). Model selection and multimodel inference: a practical information-theoretic approach. Springer.} 




\bibitem{CMS:2018yxg}
A.~M.~Sirunyan \textit{et al.} [CMS],
Phys. Lett. B \textbf{792}, 345-368 (2019)

\bibitem{ATLAS:2023vxg}
G.~Aad \textit{et al.} [ATLAS],
JHEP \textbf{07}, 090 (2023)

\bibitem{ATLAS:2024uvu}
G.~Aad \textit{et al.} [ATLAS],
Phys. Rev. D \textbf{110}, no.7, 072008 (2024)

\bibitem{BaBar:2016sci}
J.~P.~Lees \textit{et al.} [BaBar],
Phys. Rev. D \textbf{94}, no.1, 011102 (2016)

\bibitem{Bellini:2011rx}
G.~Bellini, J.~Benziger, D.~Bick, S.~Bonetti, G.~Bonfini, M.~Buizza Avanzini, B.~Caccianiga, L.~Cadonati, F.~Calaprice and C.~Carraro, \textit{et al.}
Phys. Rev. Lett. \textbf{107}, 141302 (2011)




\bibitem{Harnik:2012ni}
R.~Harnik, J.~Kopp and P.~A.~N.~Machado,
JCAP \textbf{07}, 026 (2012)

\bibitem{NA64:2024klw}
Y.~M.~Andreev \textit{et al.} [NA64],
Phys. Rev. Lett. \textbf{132}, no.21, 211803 (2024)


\bibitem{Escudero:2019gzq}
M.~Escudero, D.~Hooper, G.~Krnjaic and M.~Pierre,
JHEP \textbf{03}, 071 (2019)


\bibitem{Ghosh:2024cxi}
D.~K.~Ghosh, P.~Ghosh, S.~Jeesun and R.~Srivastava,
Phys. Rev. D \textbf{110}, no.7, 075032 (2024)

\bibitem{CCFR:1991lpl}
S.~R.~Mishra \textit{et al.} [CCFR],
Phys. Rev. Lett. \textbf{66}, 3117-3120 (1991)








 
\end{thebibliography}
\end{document}